\documentclass[11pt,a4paper]{article}

\usepackage{afterpage}
\usepackage[]{algorithm2e}
\usepackage{alltt}
\usepackage{amsfonts}
\usepackage[centertags]{amsmath}
\usepackage{amssymb}
\usepackage{amsthm}
\usepackage{array}
\usepackage{bm}
\usepackage[font=small,skip=5pt]{caption}
\usepackage{color}
\usepackage{graphicx}
\usepackage[hidelinks]{hyperref}
\usepackage{natbib}
\usepackage{subfig}
\usepackage{ulem}
\usepackage{url}
\usepackage{verbatim}
\usepackage[dvipsnames,svgnames*]{xcolor}

\SetEndCharOfAlgoLine{}
\def\app#1#2{%
  \mathrel{%
    \setbox0=\hbox{$#1\sim$}%
    \setbox2=\hbox{%
      \rlap{\hbox{$#1\propto$}}%
      \lower1.1\ht0\box0%
    }
    \raise0.25\ht2\box2%
  }%
}

\title{Updating Variational Bayes:\\  Fast sequential posterior inference}
\author{Nathaniel Tomasetti\thanks{Department of Econometrics and Business Statistics, Monash University, VIC, 3800, Australia. Email: \textit{nathaniel.tomasetti@monash.edu}. Code available at \url{https://github.com/NTomasetti/UVB_Code}}
\and Catherine Forbes\thanks{Corresponding author. Department of Econometrics and Business Statistics, Monash University, VIC, 3800, Australia. Email: \textit{catherine.forbes@monash.edu}. This author acknowledges financial support under the Australian Research Council Discovery Grant No. DP150101728.}  
\and Anastasios Panagiotelis\thanks{Department of Econometrics and Business Statistics, Monash University, VIC, 3800, Australia. Email: \textit{anastasios.panagiotelis@monash.edu}.}}

\begin{document}
\maketitle

\begin{abstract}
Variational Bayesian (VB) methods produce posterior inference in a time frame considerably smaller than traditional Markov Chain Monte Carlo approaches. Although the VB posterior is an approximation, it has been shown to produce good parameter estimates and predicted values when a rich classes of approximating distributions are considered. In this paper we propose Updating VB (UVB), a recursive algorithm used to update a sequence of VB posterior approximations in an online setting, with the computation of each posterior update requiring only the data observed since the previous update. An extension to the proposed algorithm, named UVB-IS, allows the user to trade accuracy for a substantial increase in computational speed through the use of importance sampling. The two methods and their properties are detailed in two separate simulation studies. Two empirical illustrations of the proposed UVB methods are provided, including one where a Dirichlet Process Mixture model with a novel posterior dependence structure is repeatedly updated in the context of predicting the future behaviour of vehicles on a stretch of the US Highway 101.\\


\emph{\bigskip}


\end{abstract}

\section{Introduction}
\label{sec:intro}

Time series data often arrives in high frequency streams in applications that may require a response within a very short period of time. For example, self-driving vehicles may need to constantly monitor the position of each surrounding vehicle, predict or infer the behaviour of their human drivers, and react accordingly. In this context, the most recently received data can be highly informative for very short-term predictions, if the inferred models can be processed very quickly in an online fashion. In order to account for uncertainty in the models or predictions, Bayesian updating methods may be employed by targeting a sequence of posterior distributions, each conditioned on an expanding dataset. The computational demands of such an algorithm may be improved if the incorporation of additional data does not require the re-use of any observations that have previously been conditioned upon.
\\

In many empirical settings, the desired Bayesian posterior distributions are not analytically tractable. In such cases posterior inference may be obtained using Markov chain Monte Carlo (MCMC) methods, which result in a (dependent) sample from the posterior. Unfortunately, this approach typically involves relatively slow algorithms that are incompatible with the time frames demanded by streaming data, as the computation of each posterior update involves the entire currently observed dataset. Further, while particle filtering methods for sequential posterior updating have been developed both for static parameter models (\citealp{Chopin2002}) and dynamic latent variable models (e.g. \citealp{Doucet2001}), these available methods appear to be too slow for practical online use.  This is particularly the case when they require use of the entire dataset to avoid particle degeneracy and/or when the number of inferred parameters is large. For a recent review of particle filtering methods, see \cite{Doucet2018}. An alternative approach appears in \cite{Jasra2010} and \cite{DelMoral2015}, who apply Approximate Bayesian Computation (ABC) for sequential posterior updating, however this involves an embedded particle filter that similarly scales poorly to higher dimensional models. \cite{Bhattacharya2018} provide a sequential method to update parameter inference, however their grid-based posterior evaluation is suitable only for low dimensions. Taking a different approach, \cite{Chen2019} learn the parameters of a so-called flow operator, a neural network that approximates a function which maps a set of particles from a posterior distribution at one time period, and additional data, to a set of particles belonging to an updated posterior distribution.
\\

An alternative approach that has grown in popularity in the recent literature for high dimensional models is the so-called Variational Bayes (VB) method \citep[see][for a review]{Zhang2017}. VB approximates the posterior with a tractable family of  distributions, and chooses a member of this family by minimising a particular loss function with respect to auxiliary parameters. Early work in VB found an optimal approximation with coordinate descent algorithms in exponential family models, an approach widely known as Mean Field Variational Bayes (MFVB, see \citealp{Jordan1999},  \citealp{Attias1999} \citealp{Ghahramani2000a}, \citealp{Wainwright2008}). Recent developments in VB consider gradient based algorithms (\citealp{Ranganath2014}, \citealp{Kingma2014a}), which allow for a much richer class of models and approximating distributions to be utilised. These gradient based approaches are stochastic, and target the true gradient of a given loss function with an unbiased estimator. We refer to this approach as Stochastic Variational Bayes (SVB).
\\

There is a rich tradition of using only a subset of the complete dataset for certain aspects of VB inference, such as for gradient estimation. \cite{Hoffman2010} and \cite{Wang2011} propose MFVB algorithms for Dirichlet Process Mixture (DPM) models where the optimisation of a subset of the auxiliary parameter vector occurs through gradient based approaches, using a subsample of the complete data at each iteration. \cite{Hoffman2013} and \cite{Titsias2014} implement this data subsampling into the fully gradient based SVB approaches. Alternatively, \cite{Sato2001} considers an alternative loss function defined as the expected value of the Kullback-Leibler (KL) divergence, with respect to the data generating process. Any realisation from the data generating process may be used within the MFVB coordinate descent algorithm, which is applied online with newly observed data substituted in as it becomes available. However, each of these approaches results in only a single posterior distribution conditioned on data up to some pre-specified time period $T_n$, and do not provide a mechanism for the approximation to be updated at a later time period $T_{n+1}$ following the availability of additional observations.
\\

\cite{Smidl2004} and \cite{Broderick2013} each consider VB approximations for Bayesian updating, resulting in a progressive sequence of approximate posterior distributions that each condition on data up to any given time period $T_n$. Their approaches update to the time $T_{n+1}$ by substitution of the time $T_n$ posterior with  MFVB approximations, which are feasibly obtained due to assuming the model and approximation each adhere to a suitably defined exponential family form.  In these special settings, MFVB is able to linearly combine the available optimally converged auxiliary parameters. While \cite{Smidl2004} is concerned with state space models, \cite{Broderick2013} considers application to a latent Dirichlet allocation problem, and shows it performs favourably compared to the approach of \cite{Hoffman2010} in terms of log predictive score and computational time.
\\

In this paper we formalise and extend the prior approximation approach, developing a new algorithm that we call Updating Variational Bayes (UVB).  UVB can be applied to sequentially update posterior distributions, and in a manner suitable for applications of streaming data. UVB treats data as arriving in a sequence, with the production of recursive, but approximate, posterior distributions obtained from conditioning on past information at nominated time points according to a Bayesian updating scheme. The approach delivers the approximate posterior distributions to the user at any desired point in time, with each new update using only the data observed since the previous update time. UVB requires an optimisation step for each update, which may be too slow for practical use. To reduce the computational load of repeated updates, we extend UVB to a second algorithm, called Updating Variational Bayes with Importance Samping (UVB-IS). Significant gains in computation speed per update can be achieved, albeit with some potential cost in gradient estimator variance and subsequently accuracy.  Our proposed UVB-IS shares some similarities with the gradient estimator of \cite{Sakaya2017}, however the important distinction is that our proposed UVB-IS is developed for the sequential updating setting.
\\

We provide two simulation studies: a small scale time series forecasting application, and a larger clustering application, to compare the approximation error of each of SVB, UVB, and UVB-IS relative to exact inference obtained using MCMC. We also compare the computational time required by each of the variational approximations, and show that UVB-IS is substantially faster than either UVB or SVB, while incurring only a minor cost in performance, dependent on the application. We also demonstrate the application of UVB and UVB-IS to a simple hierarchical model to re-analyse the eight schools problem of \cite{Gelman1997}, and measure the increased approximation error from the updating approaches relative to SVB in this setting.
\\

Finally we demonstrate the application of UVB to the problem of updating posterior inference in the context of a DPM model. Here the aim is to provide Bayesian inference and prediction regarding the heterogeneous behaviour of 500 drivers from the New Generation Simulation dataset \citep{NGSIM2017}, according to the distribution of their lateral lane position. In this context, data arrives rapidly. We introduce a new class of dependent approximating distributions, and show that the DPM model with UVB based inference is able to provide more accurate forecasts than those achieved using a standard MFVB based approach. UVB in this case has accuracy comparable to repeated use of (full data) SVB, but benefits from an ability to process updates sequentially as additional data arrives.
\\ 
 
The paper is arranged as follows: in Section~\ref{sec:background} we review standard VB methods and the available gradient algorithms commonly employed. In Section~\ref{sec:UVB} we propose our main UVB approach, with the UVB-IS extension detailed in Section~\ref{sec:UVBIS}. Next, Section~\ref{sec:UVBSim} contains simulation studies for time series data and a mixture distribution, while Section~\ref{sec:eightschools} details applications of the newly proposed methods to the Eight Schools hierarchical model of \cite{Gelman2014}. UVB is applied to a vehicle DPM model in Section~\ref{sec:lanepos}, and Section~\ref{sec:UVBSummary} concludes the paper.

\section{Background on Variational Bayes}
\label{sec:background}
Before introducing our novel algorithms for recursively updating approximations to the posterior, the main ideas associated with the implementation of an SVB approach are introduced. A more detailed description of SVB can be found in \cite{Blei2017}, with further references provided therein.  
\\

The usual target of Bayesian inference is the posterior distribution for a potentially vector-valued static parameter ${\bm\theta}$, as characterised by its probability density function (pdf) denoted by $p({\bm\theta} | {\bm y}_{1:T})$. Here ${\bm y}_{1:T}$ denotes data observed from time $1$ to $T$ and the posterior pdf is obtained using Bayes' theorem, given by
\begin{equation}
\label{posterior}
p({\bm\theta} | {\bm y}_{1:T}) = \frac{p({\bm y}_{1:T} | {\bm\theta})p({\bm\theta})}{\int_{\bm \theta}p({\bm y}_{1:T} | {\bm\theta})p({\bm\theta})d{\bm\theta}}\,,
\end{equation}
where $p({\bm\theta})$ denotes the pdf for the prior distribution that characterises belief about ${\bm\theta}$ prior to the observation of ${\bm y_{1:T}}$. Although MCMC algorithms are commonly used to produce a (typically dependent) sample from this posterior distribution, these can be computationally intensive.  
\\

As an alternative to MCMC, VB aims to approximate the pdf in (\ref{posterior}) with another density of given parametric form, denoted by $q_{\bm \lambda}({\bm\theta} |{\bm y}_{1:T})$.  Here ${\bm \lambda}$ is a vector of auxiliary parameters associated with the approximation, to be selected via optimisation. We note that the distribution $q$ is explicitly parameterised by $\bm \lambda$, and any evaluation of this distribution does not require $\bm y_{1:T}$. The approximating distribution depends on ${\bm y}_{1:T}$ only through the choice of $\bm{\lambda}$, however it is included in the notation to reinforce that $q_{\bm \lambda}({\bm\theta} |{\bm y}_{1:T})$ is an approximation to the posterior distribution $p(\bm{\theta} | {\bm y}_{1:T})$.
\\

In the SVB context, the family of the approximating distribution $q_{\bm \lambda}$ is held fixed, with the member of that family indexed by the parameter vector ${\bm \lambda}$ selected to minimise a given loss function. Typically the KL divergence  \citep{Kullback1951} from $q_{\bm \lambda}({\bm \theta} |{\bm y}_{1:T})$ to $p({\bm\theta} |{\bm y}_{1:T})$, denoted as $KL[q_{\bm \lambda}({\bm\theta} |{\bm y}_{1:T})\hspace{.1cm}||\hspace{.1cm}p({\bm\theta} |{\bm y}_{1:T})]$, is used, with
\begin{equation}
\label{KL-def}
KL[q_{\bm\lambda}({\bm\theta} |{\bm y}_{1:T})\hspace{.1cm}||\hspace{.1cm}p({\bm\theta} |{\bm y}_{1:T})] = E_q \left[ \log(q_{\bm \lambda}({\bm \theta} |{\bm y}_{1:T})) - \log(p({\bm \theta} |{\bm y}_{1:T})) \right]\,.
\end{equation}

Often in practice, the KL divergence in (\ref{KL-def}) is intractable, with $p({\bm\theta} | {\bm y}_{1:T})$ only known up to a proportionality constant due to the difficulties involved in the evaluation of the integral in the denominator of (\ref{posterior}). Nevertheless, it has been shown that an equivalent problem to minimising the KL divergence is to maximise the so-called \textit{evidence lower bound} (ELBO \citealp{Attias1999}), given by
\begin{equation}
\label{ELBO}
\mathcal{L}(q, {\bm\lambda}) = E_q \left[\log(p({\bm \theta}, {\bm y}_{1:T})) - \log(q_{\bm\lambda}({\bm\theta} | {\bm y}_{1:T}))\right] .
\end{equation}

A further complication that typically arises when attempting to implement SVB is that an analytical expression for the expectation in (\ref{ELBO}) may not be available. In this case, maximisation of the ELBO may be achieved via stochastic gradient ascent (SGA, \citealp{Bottou2010}). To apply SGA to the problem of maximising the ELBO, an inital value ${\bm \lambda}^{(1)}$ is selected and is recursively modified to $\bm{\lambda}^{(m)}$, for $m = 2, 3, \ldots,$ according to
\begin{equation}
\label{gradientAscent}
{\bm\lambda}^{(m+1)} = {\bm\lambda}^{(m)} + \rho^{(m)} \widehat{\frac{\partial\mathcal{L}(q, {\bm\lambda})}{\partial {\bm\lambda}}} \bigg\rvert_{{\bm\lambda} = {\bm\lambda}^{(m)}}\ 
\end{equation}
with the final value of ${\bm\lambda}^{(M)}$ obtained when the change from $\mathcal{L}(q, {\bm\lambda}^{(M-1)})$ to $\mathcal{L}(q, {\bm \lambda}^{(M)})$ falls below some pre-specified threshold \citep{Hoffman2013}.
\\

The adjustment term in (\ref{gradientAscent}) is made of two factors, the so-called \textit{learning rate}, $\rho^{(m)}$, and an estimate of the gradient of the ELBO, $\widehat{\frac{\partial\mathcal{L}(q, {\bm\lambda})}{\partial {\bm\lambda}}}$. A popular estimator of this gradient is the score-based estimator \citep{Ranganath2014}, given by
\begin{equation}
\label{scoreDeriv} 
\widehat{\frac{\partial\mathcal{L}(q, {\bm\lambda})}{\partial {\bm\lambda}}}_{SC} = \frac{1}{S}\sum_{j = 1}^S \frac{\partial \log(q_{\bm \lambda}({\bm\theta}^{(j)} | {\bm y}_{1:T}))}{\partial {\bm\lambda}} \left(\log(p(\bm y_{1:T}, {\bm\theta}^{(j)})) - \log(q_{\bm\lambda}({\bm\theta}^{(j)} | {\bm y}_{1:T})) - \widehat{\bm a} \right),
\end{equation}
where the simulated values $\{ \bm{\theta}^{(j)}, \mbox{ for } j = 1, 2, \ldots, S\}$ are drawn from the presiding approximating density $q_{\bm\lambda^{(m)}}({\bm\theta} | {\bm y}_{1:T})$, and $\widehat{\bm a}$ is a vector of control variates with
\begin{equation}
\label{controlVar}
\widehat{a}_k = \frac{\widehat{\mbox{Cov}}\left(\frac{\partial \log(q_{\bm \lambda}({\bm\theta} | {\bm y}_{1:T}))}{\partial {\bm\lambda}_k} \left(\log(p({\bm y}_{1:T}, {\bm\theta} )) - \log(q_{\bm\lambda}({\bm\theta} | {\bm y}_{1:T}))\right), \frac{\partial \log(q_{\bm \lambda}({\bm\theta} | {\bm y}_{1:T}))}{\partial {\bm\lambda}_k}\right)}{\widehat{\mbox{Var}}\left(\frac{\partial \log(q_{\bm \lambda}({\bm\theta} | {\bm y}_{1:T}))}{\partial {\bm\lambda}_k}\right)}.
\end{equation}
As (\ref{scoreDeriv}) results in an unbiased estimator of the gradient of the ELBO, it is known  that the SGA procedure will converge in probability to a local maximum \citep{Robbins1951}, provided that the learning rate sequence\footnote{The learning rate used for all implementations of SGA in this paper is provided by the Adaptive Moment (Adam) algorithm of \citet{Kingma2014b}.} satisfies
\begin{equation}
\sum_{m=1}^{\infty} \rho^{(m)} =  \infty 
\end{equation}
and
\begin{equation}
\sum_{m=1}^{\infty} (\rho^{(m)})^2 <  \infty.
\end{equation}
\\
We note that although SGA is itself a recursive procedure, the result in the VB context is the one-time posterior pdf approximation $q_{\bm{\lambda}^{*}}\approx p({\bm\theta} | {\bm y}_{1:T})$, where $\bm \lambda^* = \bm \lambda^{(M)}$ is the optimal parameter.
\\

\subsection{Dependence in the Approximation}

Considering the vector $\bm \theta = (\theta_1, \theta_2)^{\prime}$, the application of SVB often employs the so-called Mean Field approximation \citep{Bishop2006} where the approximating distribution is factorised as
\begin{equation}
\label{VB:factorised}
q_{\bm \lambda}(\theta_1, \theta_2 | \bm y_{1:T}) = q_{\bm \lambda}(\theta_1 | \bm y_{1:T})q_{\bm \lambda}(\theta_2 | \bm y_{1:T}).
\end{equation}
However SVB allows more general forms of the approximating distribution that may include dependence, for example
\begin{equation}
\label{VB:dependence}
q_{\bm \lambda}(\theta_1, \theta_2 | \bm y_{1:T}) = q_{\bm \lambda}(\theta_1 | \bm y_{1:T})q_{\bm \lambda}(\theta_2 | \theta_1, \bm y_{1:T}).
\end{equation}
In this paper we also consider approximating distribution families that include the true posterior distribution for the subset $\theta_2$, when available. We implicitly use the marginal posterior for the approximation in Section 5, while in Section 7 we also explicitly include the exact conditional distribution in place of the approximation, i.e. 
\begin{equation}
\label{VB:dependenceExact}
q_{\bm \lambda}(\theta_1, \theta_2 | \bm y_{1:T}) = q_{\bm \lambda}(\theta_1 | \bm y_{1:T})p(\theta_2 | \theta_1, \bm y_{1:T}).
\end{equation}
To our knowledge, the potential to exploit this conditional approximation structure - and in particular to include an exact component within that structure - appears to be a novel contribution to the literature.

\section{Updating Variational Bayes}
\label{sec:UVB}

We now introduce the proposed algorithm for updating VB when data is observed in an online setting.  Let $T_1, T_2, \ldots$ be a sequence of time points, from which a sequence of posterior distributions $p({\bm\theta} | {\bm y}_{1:T_1}), p({\bm\theta} | {\bm y}_{1:T_2}), \ldots$, is desired. Now suppose that the (exact) posterior distribution for the governing (static) parameter vector $\bm{\theta}$ is available, as given by its pdf $p({\bm\theta} | {\bm y}_{1:T_n})$. Our objective is to update this posterior distribution, after observing data up to, and including, time $T_{n+1}$, when the additional $T_{n+1} - T_n$ data points have become available. The pdf of the resulting updated posterior distribution is denoted as $p({\bm\theta} | {\bm y}_{1:T_{n+1}})$. In an online setting, where new data continues to appear, we will want to repeat this updating procedure sequentially, each time updating the past posterior to reflect all of the data, including the latest available.
\\

The usual application of Bayes' rule at a given time $T_{n+1}$ involves a likelihood made up of $T_{n+1}$ factors.  However with the availability of the posterior at time $T_n$, given by its density $p({\bm\theta} | {\bm y}_{1:T_{n}})$, the updated time $T_{n+1}$ posterior is given by
\begin{equation}
\label{update:updatePost}
p({\bm\theta} | {\bm y}_{1:T_{n+1}}) \propto p({\bm y}_{T_{n}+1:T_{n+1}} |  {\bm y}_{1:T_{n}}, {\bm\theta})p({\bm\theta} | {\bm y}_{1:T_{n}}),
\end{equation}
where $p({\bm y}_{T_{n}+1:T_{n+1}} | {\bm\theta}, {\bm y}_{1:T_{n}})$ on the right hand side of (\ref{update:updatePost}) is comprised of only $T_{n+1}-T_n$ factors.
\\

We propose the Updating Variational Bayes (UVB) algorithm  for use when the evaluation of the online posterior updating is computationally demanding. Our UVB algorithm, detailed in Algorithm \ref{alg:UVB}, is initialised by forming the variational approximation at a given time $T_1$ as $q_{\bm \lambda_1^*}(\bm \theta | \bm y_{1:T_1})$ where
\begin{equation}
\label{UVB:Time1}
\bm \lambda_1^* = \arg \underset{\bm \lambda_1}{\min} \hspace{1mm} KL[q_{\bm \lambda_1}(\bm \theta | \bm y_{1:T_1}) \hspace{.1cm}||\hspace{.1cm}p(\bm \theta | \bm y_{1:T_1})].
\end{equation}
At this first stage, we simply approximate the first posterior $p(\bm \theta | \bm y_{1:T_1})$ with the optimised distribution as in SVB, namely $q_{\bm \lambda_1^*}(\bm \theta | \bm y_{1:T_1}).$  
\\

In general then, after approximating the posterior at time $T_n$ with $q_{\bm \lambda_{n}^*}(\bm \theta | \bm y_{1:T_{n}})$ and observing additional data up to time $T_{n+1}$, UVB replaces the posterior construction described by (\ref{update:updatePost}) with the available approximation,
\begin{equation}
\label{UVB:altPosterior}
\widetilde{p}(\bm \theta |  \bm y_{1:T_{n+1}}) \propto p(\bm y_{T_{n}+1:T_{n+1}} | \bm \theta)q_{\bm \lambda_{n}^*}(\bm \theta | \bm y_{1:T_{n}}).
\end{equation}
This defines an alternate target distribution, $\widetilde{p}(\bm \theta |  \bm y_{1:T_{n+1}}),$ referred to as the `pseudo-posterior' at time $T_{n+1}$.
\\

The objective for each update is to find ${\bm \lambda_{n+1}^*}$ (and hence $q_{\bm \lambda_{n+1}^*}(\bm \theta | \bm y_{1:T_{n+1}}))$ through the minimisation of the KL divergence to the corresponding pseudo-posterior, resulting in
\begin{equation}
\label{UVB:TimeNp1}
\bm \lambda_{n+1}^* = \arg \underset{\bm \lambda_{n+1}}{\min} \hspace{1mm} KL[q_{\bm \lambda_{n+1}}(\bm \theta | \bm y_{1:T_{n+1}})  \hspace{.1cm}||\hspace{.1cm}\widetilde{p}(\bm \theta | \bm y_{1:T_{n+1}})],
\end{equation}
for each $n=1,2, \ldots$. The sequence of distributional families $q_{\bm \lambda_1}, q_{\bm \lambda_2}, \ldots,$ may differ at each time period, though we note it is convenient to hold the family fixed. 
\\

\RestyleAlgo{boxruled}
\begin{algorithm}[htbp]
 \SetKwInOut{Input}{Input}
 \Input{Prior, Likelihood.}
 \KwResult{Posterior approximation at $T_{\tau}$.}
 Observe $\bm y_{1:T_1}$.\;
 Minimises $KL[q_{\bm\lambda_1}(\bm \theta |\bm y_{1:T_1})\hspace{.1cm}||\hspace{.1cm}p(\bm \theta |\bm y_{1:T_1})]$ using SGA via (\ref{scoreDeriv}).\;
 \For{$n \mbox{ in } 1, \ldots, \tau -1$}{
   Observe next data $\bm y_{T_{n}+1:T_{n+1}}$.\;
   Use $q_{\bm\lambda_{n}}(\bm \theta | \bm y_{1:T_{n}})$ and (\ref{UVB:altPosterior}) to construct the UVB pseudo-posterior up to proportionality.\;
   Minimise $KL[q_{\bm\lambda_{n+1}}(\bm \theta |\bm y_{1:T_{n+1}}) \hspace{.1cm}||\hspace{.1cm}\widetilde{p}(\bm \theta |\bm y_{1:T_{n+1}})]$ using SGA via (\ref{scoreDeriv}).\;
  }
 \caption{Updating Variational Bayes (UVB)}
  \label{alg:UVB}
\end{algorithm}

We note some important features of the proposed UVB algorithm compared with an SVB implementation. First, at time $T_{n+1}$ an SVB implementation would target the exact posterior $p(\bm \theta | \bm y_{1:T_{n+1}}) \propto p(\bm y_{1:T_{n+1}} | \bm \theta)p(\bm \theta)$ whereas UVB instead targets an alternate pseudo-posterior distribution in (\ref{UVB:altPosterior}). Second, at time $T_{n+1}$, the evaluation for UVB corresponding to (\ref{scoreDeriv}) is composed of only $T_{n+1}-T_{n}$ factors. Hence, the computational complexity of UVB has rate $O(T_{n+1}-T_{n})$ rather than rate $O(T_{n+1})$, i.e. computing UVB is not increasing in the number of observations for equally spaced intervals, as is the case for SVB. Third, unlike SVB, the UVB algorithm can begin even when only part of the data has been observed, making it well-suited to online applications. Further, the prevailing optimal value of $\bm\lambda_n$, denoted $\bm \lambda_n^*$, could be used as the UVB starting value for the optimisation at time $T_{n+1}$ as long as the class of approximating distributions $q$ is the same for each update.  This may reduce the number of SGA iterations required for the UVB algorithm to converge. However, even with frequent updating via UVB, there will be some loss of accuracy in the posterior approximation at time $T_{n+1}$, $q_{\bm \lambda_{n+1}}(\bm \theta | \bm y_{1:T_{n+1}})$, relative to the corresponding approximation obtained by SVB.  The extent of this loss will be context specific.
\\

Clearly there will be a trade-off between computational speed and accuracy. These issues are investigated in a simulation setting in Section \ref{sec:UVBSim}. Before exploring these aspects, we introduce a modified approach whereby the computational speed may be further improved, albeit potentially with some additional loss in accuracy. This modified approach, referred to as UVB with Importance Sampling (UVB-IS), is described in the next section.

\section{UVB with Importance Sampling} 
\label{sec:UVBIS}

An application of UVB up to some time $T_n$ involves SVB inference at time $T_1$ followed by $n-1$ updates, for a total of $n$ applications of SGA optimisation. Repeated updates may incur a significant computational overhead relative to SVB, which applies only a single SGA algorithm using all data up to time $T_n$. In this section we address this problem and explore the possibility of achieving large computational gains per update through the incorporation of ideas from importance sampling. (For a general overview of importance sampling, see \cite{Gelman2014}.) Before introducing our UVB with Importance Sampling (UVB-IS) algorithm, we briefly review the incorporation of importance sampling into SGA, as introduced by \cite{Sakaya2017}.
\\ 

Temporarily suppressing the subscript $n$ on the given time period $T$, the $m^{th}$ iteration in the SGA algorithm for a given target VB posterior changes $\bm\lambda^{(m)}$ to $\bm\lambda^{(m+1)}$ via 
$S$ simulations of $\bm \theta^{(m)}$ from $q_{\bm \lambda^{(m)}}(\bm{\theta} | \bm y_{1:T})$ as per (\ref{scoreDeriv}). For each of these simulations, the log-likelihood, log-prior, and additional terms involving $q_{\bm \lambda^{(m)}}(\bm \theta | \bm y_{1:T})$ must be evaluated. Note that, for large scale applications this computation is dominated by the $T$ terms in the log-likelihood. 
\\ 

In the subsequent SGA iteration from $\bm\lambda^{(m+1)}$ to $\bm\lambda^{(m+2)}$, the evaluation of the log-likelihood requires a new set of $S$ simulations $\bm \theta^{(m+1)}$ from $q_{\bm\lambda^{(m+1)}}(\bm{\theta} |\bm y_{1:T})$. \cite{Sakaya2017} note that as the change from $\bm\lambda^{(m)}$ to $\bm\lambda^{(m+1)}$ is likely to be small,
the distributions $q_{\bm\lambda^{(m)}}(\bm{\theta} |\bm y_{1:T})$ and $q_{\bm\lambda^{(m+1)}}(\bm{\theta} |\bm y_{1:T})$ will likely be similar. Using this motivation, an alternative gradient estimator is suggested for each iteration $k = m+1, m+2, \ldots, m+r$ via an importance sampler that uses $q_{\bm\lambda^{(m)}}(\bm{\theta} | \bm y_{1:T})$ as a proposal distribution, rather than generating new draws of $\bm{\theta}$ from each $q_{\bm\lambda^{(k)}}(\bm{\theta} | \bm y_{1:T})$. This approach retains the set of samples $\bm \theta^{(m)}$ and their associated log-likelihood values, only resampling $\bm \theta$ and re-evaluating the corresponding log-likelihood at iteration $m+r+1$.  In the SVB context the value of $r$ should not be taken to be too large, as substantial differences between $\bm\lambda^{(m)}$ and $\bm\lambda^{(m+r)}$ may lead to a corresponding increase in the variance of the resulting gradient estimator.
\\

In the context of UVB, we sequentially update the posterior approximation at each time $T_n$ via repeated applications of SGA. As before UVB-IS holds the family of the approximating distribution $q_{\bm \lambda}$ fixed between each update, and sets the initial value of the parameter vector at time $T_{n+1}$ equal to the optimal value from the previous update, i.e. we set $\bm\lambda_{n+1}^{(1)} = \bm\lambda_{n}^*$. During the subsequent application of SGA, the sequence of parameter vectors $\bm\lambda_{n+1}^{(1)}, \bm\lambda_{n+1}^{(2)}, \ldots, \bm\lambda_{n+1}^*$ corresponds to a sequence of distributions moving from $q_{\bm\lambda_{n}^*}(\bm{\theta} | \bm y_{1:T_n})$ to $q_{\bm\lambda_{n+1}^*}(\bm{\theta} | \bm y_{1:T_{n+1}})$. For repeated updates with small values of $T_{n+1}-T_n$, the new information about $\bm{\theta}$ in $\bm y_{T_n+1:T_{n+1}}$ will typically be relatively small, and unless there is a structural change in the data process, we expect the approximating distributions will become similar. 
\\

The above observation motivates the addition of an importance sampling gradient estimator to be applied for each update. In each update using the SGA algorithm at time $T_{n+1}$ all of the requisite gradients are estimated via importance sampling, using the previous UVB posterior $q_{\bm\lambda_{n}^*}(\bm{\theta} | \bm y_{1:T_n})$ as the (identical) proposal distribution. The consequence of this approach is that only $S$ $\bm {\theta}$ samples are required for the entire SGA algorithm, and the likelihood is evaluated $S$ times in total, rather than $S$ times per iteration (or $S$ times per $r$ iterations in the case of \cite{Sakaya2017}). 
\\

Suppressing the SGA iteration superscript index $(m)$, the UVB-IS gradent estimator is derived from the score-based estimator implied by (\ref{scoreDeriv}). In this case, the updated joint distribution, given by $p(\bm y_{T_n+1:T_{n+1}}, \bm \theta |\bm y_{1:T_{n}})$, is replaced by an expression proportional to (\ref{UVB:altPosterior}), with
\begin{equation}
\label{UVBIS:scoreGrad}
\frac{\partial\mathcal{L}(q, {\bm\lambda}_{n+1})}{\partial {\bm \lambda}_{n+1}} =
 \int_{\bm\theta} q_{{\bm\lambda}_{n+1}}({\bm\theta} | {\bm y}_{1:T_{n+1}}) \underbrace{\frac{\partial \log(q_{{\bm\lambda}_{n+1}} | {\bm y}_{1:T_{n+1}})}{\partial {\bm\lambda}_{n+1}} \left(\log \left(\frac{\widetilde{p}(\bm y_{T_n+1:T_{n+1}}, \bm \theta |\bm y_{1:T_{n}})}{q_{{\bm\lambda}_{n+1}}({\bm \theta | {\bm y}_{1:T_{n+1}})}} \right) - \widehat{\bm a} \right)}_{f(\bm \theta)} d{\bm\theta}.
\end{equation}
Multiplication and division of the integrand in (\ref{UVBIS:scoreGrad}) by $q_{{\bm\lambda}_n^*}({\bm\theta} | {\bm y}_{1:T_n})$ allows it to be written as an expectation with respect to $q_{{\bm\lambda}_n^*}({\bm\theta} | {\bm y}_{1:T_n})$,
\begin{equation}
\label{UVBIS:scoreGradIS}
\frac{\partial\mathcal{L}(q, {\bm \lambda}_{n+1})}{\partial {\bm \lambda}_{n+1}} = \int_{\bm\theta} q_{{\bm \lambda}_{n}^*}({\bm \theta} | {\bm y}_{1:T_{n}})\frac{q_{{\bm\lambda}_{n+1}}({\bm\theta} | {\bm y}_{1:T_{n+1}})}{q_{{\bm\lambda}_{n}^*}(\bm{\theta} | {\bm y}_{1:T_{n}})} f(\bm \theta)d{\bm \theta},
\end{equation}
where we note the factor $f(\bm \theta)$ is defined using the underbrace in (\ref{UVBIS:scoreGrad}). Hence, (\ref{UVBIS:scoreGradIS}) may be estimated via a Monte Carlo average,
\begin{equation}
\label{UVBIS:scoreEstIS}
\widehat{\frac{\partial\mathcal{L}(q, {\bm \lambda}_{n+1})}{\partial {\bm\lambda}_{n+1}}}_{IS} = \frac{1}{S} \sum_{j=1}^S w({\bm\theta}^{(j)})f(\bm\theta^{(j)})
\end{equation}
since ${\bm\theta}^{(j)} \sim q_{{\bm\lambda}_{n}^*}({\bm\theta} | {\bm y}_{1:{T_n}})$ and 
\begin{equation}
w({\bm\theta}^{(j)}) = \frac{q_{{\bm\lambda}_{n+1}}({\bm\theta}^{(j)} | {\bm y}_{1:T_{n+1}})}{q_{{\bm\lambda}_{n}^*}({\bm\theta}^{(j)} | {\bm y}_{1:T_{n}})},
\end{equation}
with $\widehat{\bm a}$ estimated as per Equation (\ref{controlVar}).
\\

Since only the value of ${\bm \lambda}_{n+1}$ changes in each iteration of SGA, and the $S$ samples $\bm\theta^{(j)}$ are held fixed, only the terms involving $\bm \lambda_{n+1}$, namely $\frac{\partial}{\partial {\bm\lambda}_{n+1}} \log(q_{{\bm\lambda}_{n+1}}({\bm\theta}^{(j)} | {\bm y}_{1:T_{n+1}}))$ and $q_{{\bm\lambda}_{n+1}}({\bm\theta}^{(j)} | {\bm y}_{1:T_{n+1}})$, are required to be calculated.
\\

The variance of the UVB-IS gradient estimator is increased relative to the score-based gradient estimator in (\ref{scoreDeriv}) due to the presence of the importance sampling weights. This increased variance may result in a reduction in the accuracy of $q_{{\bm\lambda}^*_{n+1}}({\bm\theta} | {\bm y}_{1:T_{n+1}})$. This is due to the fact that the algorithm stopping criterion, which is a sufficiently small value of $|\mathcal{L}(q, \bm\lambda^{(m+1)}) - \mathcal{L}(q, \bm\lambda^{(m)})|$ can only be evaluated approximately by a noisy estimator, also produced via an importance sampler. As the computation per iteration is extremely small, $S$ may be set to a larger value to reduce the variance, thereby allowing the user the capacity to balance the inevitable trade-off between computational time and approximation accuracy to suit their requirements. Provided there is no major structural change in the data generating process, it is expected that the distributions $q_{{\bm\lambda}_{n}^*}({\bm\theta} | {\bm y}_{1:T_{n}})$ and $q_{{\bm\lambda}_{n+1}^*}({\bm\theta} | {\bm y}_{1:T_{n+1}})$ become more similar as $n$ increases, subsequently reducing the UVB-IS gradient estimator variance.
\\

\begin{figure}[h]
	\subfloat[][Sakaya and Klami (2017)]{\includegraphics[width=7cm, height = 8cm]{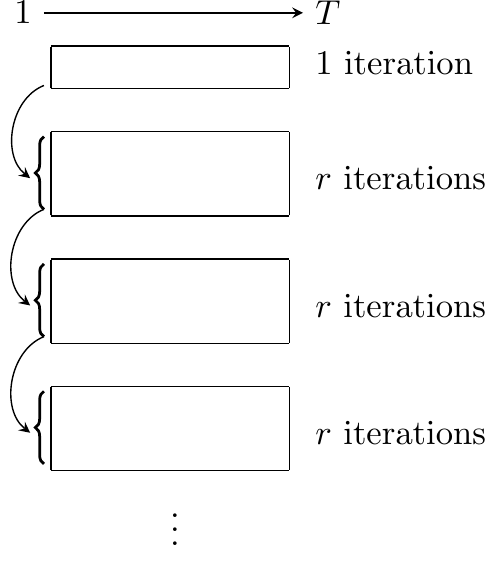}}
	\hspace{1cm}
	\subfloat[][UVB-IS]{\includegraphics[width=7cm, height = 8cm]{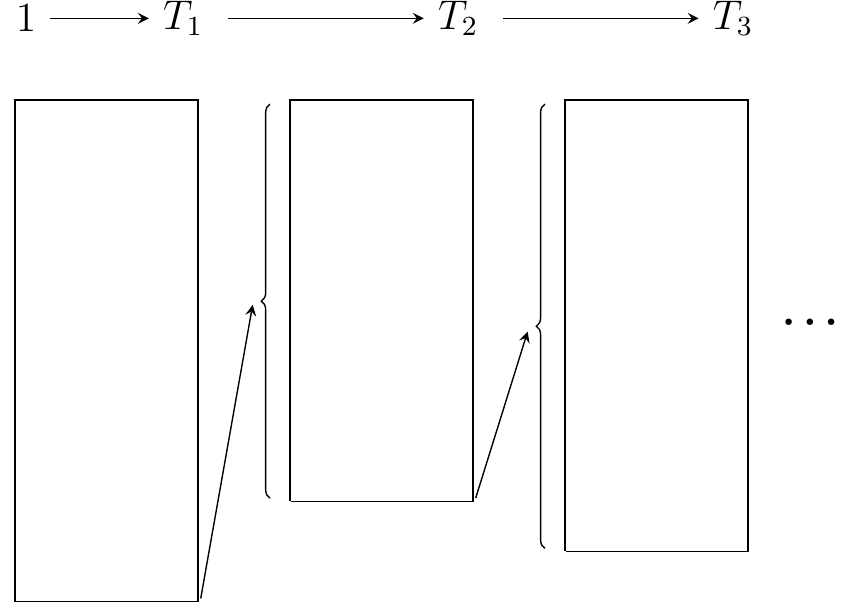}}%
	\caption{Graphical illustrations for importance sampling in SVB algorithms.
		(a): The approach of \cite{Sakaya2017}. Each block indicates $r$ iterations of a single implementation of the SGA algorithm, with arrows indicating that the final iteration of each block is used as an importance sampling proposal distribution for the next $r$ iterations contained in the subsequent block. (b): The UVB-IS algorithm, where each block indicates that SGA is applied three times, once for each of three distributional updates corresponding to an increase in data. For update, indicated by an arrow, a sample from the pseudo-posterior distribution corresponding to the previous update is used as proposal draws in every iteration of the SGA algorithm.}
	\label{fig:UVBIS}
\end{figure}

The proposed UVB-IS algorithm is summarised in Algorithm \ref{alg:UVBIS}. Figure \ref{fig:UVBIS} provides a diagram to help illustrate the differences between the approach of \cite{Sakaya2017} to UVB-IS. In panel (a) of Figure \ref{fig:UVBIS}, each block indicates $r$ separate iterations of SGA, each undertaken over an entire sample of length $T$, with arrows indicating that the final iteration of each block is used as an importance sampling proposal distribution for the entire next block. That is, there is one SGA algorithm applied for all data, but the importance sampling distribution changes every $r^{th}$ iteration until convergence is reached. In panel (b) of Figure \ref{fig:UVBIS}, three distributional updates using UVB-IS are depicted. In this case, the posterior itself is updated periodically, as indicated by arrows and corresponding to times $T_1, T_2$, and $T_3$, with the same importance sampling distribution used for \textit{all} SGA iterations needed to complete a single distributional update.\\

\vspace{2mm}
\begin{algorithm}[htbp]
	\SetKwInOut{Input}{Input}
	\Input{Prior, Likelihood.}
	\KwResult{Approximating distribution at $T_{\tau}$.}
	Observe $\bm y_{1:T_1}$.\;
 Minimises $KL[q_{\bm\lambda_1}(\bm \theta |\bm y_{1:T_1})\hspace{.1cm}||\hspace{.1cm}p(\bm \theta |\bm y_{1:T_1})]$ using SGA via (\ref{scoreDeriv}).\;
	\For{$n \mbox{ in } 1, \ldots, \tau-1$}{
		 Observe next data $\bm y_{T_{n}+1:T_{n+1}}$.\;
		Sample $\bm \theta^{(j)} \sim q_{\bm\lambda_{n}^*}(\bm \theta | \bm y_{1:T_{n}})$ for $j = 1, 2, \ldots S$.\; 
		Evaluate $p(\bm y_{T_{n}+1:T_{n+1}} |\bm \theta^{(j)})$ and $q_{\bm \lambda_{n}^*}(\bm \theta^{(j)} | \bm y_{1:T_{n}})$ for each $j$.\;
		Set $\bm \lambda_{n+1}^{(1)}$ to $\bm \lambda_{n}^*$.\;
		Minimise $KL[q_{\bm \lambda_{n+1}}(\bm \theta |\bm y_{1:T_{n+1}}) \hspace{.1cm}||\hspace{.1cm}\widetilde{p}(\bm \theta |\bm y_{1:T_{n+1}})]$ using SGA via (\ref{UVBIS:scoreEstIS}).\;
	}
	\caption{UVB with Importance Sampling (UVB-IS)}
	\label{alg:UVBIS}
\end{algorithm}

\section{Simulation Study}
\label{sec:UVBSim}

To investigate the trade-off between the computational efficiency and accuracy of different methods we consider two simulated examples. The first is a time series forecasting application, while the second is a clustering example based on a mixture model. As well as considering both of the proposed algorithms (i.e. UVB and UVB-IS) we also consider a standard SVB approach and an exact MCMC algorithm, based on a Random Walk Metropolis Hastings strategy (see \citealp{Gilks1995a}, \citealp{Gilks1995b}, and \citealp{Garthwaite2016}), employed using all data observed up to each relevant time point.

\subsection{Time Series Forecasting}
\label{subsec:UVBTS} 

In this first simulation study, we consider $R = 500$ replications of time series data, with each comprised of {T = 500} observations simulated from the following auto-regressive order 3 (AR3) model,
\begin{equation}
\label{UVB:TSAR3}
y_t = \mu + \phi_1 (y_{t-1} - \mu) + \phi_2 (y_{t-2} - \mu) + \phi_3 (y_{t-3} - \mu) + e_t
\end{equation}
where $e_t \sim N(0, \sigma^2)$. 
For each replication, the true values of the parameters are obtained by drawing $\mu$ and each auto-regressive coefficient, $\phi_1, \phi_2,$ and $\phi_3$ from an independent $N(0, 1)$ distribution, accepting only draws where each $\phi$ lies in the AR3 stationary region. The precision parameter, $\sigma^{-2}$, is drawn from a Gamma distribution with both shape and rate equal to five.
\\

The inferential objective is to progressively produce the one-step ahead predictive densities, each based on a UVB approximation to the target posterior distribution that results from assuming data arises from the AR3 model above, with a prior distribution specified for $\bm\theta = \{\mu, \phi_1, \phi_2, \phi_3, \log(\sigma^2)\}$. The prior distribution for the parameter vector is taken as $\bm \theta \sim N(\boldsymbol{0}_5, 10 \mathbb{I}_5)$, where $\boldsymbol{0}_d$ and $\mathbb{I}_d$ denote, respectively, the $d-$dimensional zero vector and identity matrix.
In particular, we aim to produce UVB-based approximate one-step ahead predictive distributions progressively, using at time $T_n$ all (and only) data up to and including time period $T_n$, recursively for each of the 21 time periods given by $T_n = 100,125,150,\ldots 500$.  That is, the first target predictive distribution is given by $p(y_{101}|\bm y_{1:100})$, followed by $p(y_{126}| \bm y_{1:125})$, and continuing on to the final predictive $p(y_{501}|\bm y_{1:500})$. For each update, predictive distributions are approximated with $q_{\bm \lambda}$ taken as a $K-$component mixture of multivariate normal distributions, with the results compared using three different choices of $K$, with $K=1, 2$ and $3$. This strategy allows us to compare the approximation accuracy of the simple $K=1$ distribution that may not adequately capture the entire posterior distribution as well as more complex approximations. 
\\

For the cases involving SVB and UVB, the  score-based gradient estimator (\ref{scoreDeriv}) uses $S = 25$ draws of ${\bm \theta}$, however we use a larger number of draws for UVB-IS to offset the increased variance, setting $S = 100$. Finally the MCMC benchmark comparison is based on 15000 posterior draws,  with the first 10000 discarded for `burn in'. In each approach we  allow $\{\phi_1, \phi_2, \phi_3\}$ to take any value in $\mathbb{R}^3$, so the posterior distribution for these parameters is not restricted to the AR3 stationary region.
\\

Under the posterior given by $p(\bm \theta | \bm y_{1:T_n})$  together with the conditional predictive densities implied by (\ref{UVB:TSAR3}), the one-step ahead predictive density is given by 
\begin{equation}
\label{UVB:TSforecastDist}
p(y_{T_n + 1} | \bm y_{1:T_n}) = \int_{\bm \theta} p(y_{T_n + 1} | \bm y_{1:T_n}, \bm \theta)p(\bm \theta | \bm y_{1:T_n})d{\bm \theta}.
\end{equation}
Given our UVB approximation to the posterior at time $T_n$, we approximate the integral in (\ref{UVB:TSforecastDist}) using $M$ draws  ${\bm\theta}^{(1)}\ldots{\bm\theta}^{(M)}\sim q_{\bm{\lambda}^*_{n}}(\bm \theta | \bm y_{1:T_n}),$ with the resulting marginal predictive density estimate given by
\begin{equation}
\label{UVB:TSforecastDistApprox}
\widehat{p}(y_{T_n + 1} | \bm y_{1:T_n}) \approx \frac{1}{M} \sum_{i=1}^M  p(y_{T_n + 1} | \bm y_{1:T_n}, \bm \theta^{(i)}).
\end{equation}

\begin{figure}[h]
\subfloat[][Forecast accuracy (MCLS)]{\includegraphics[width=7.5cm, height = 12cm]{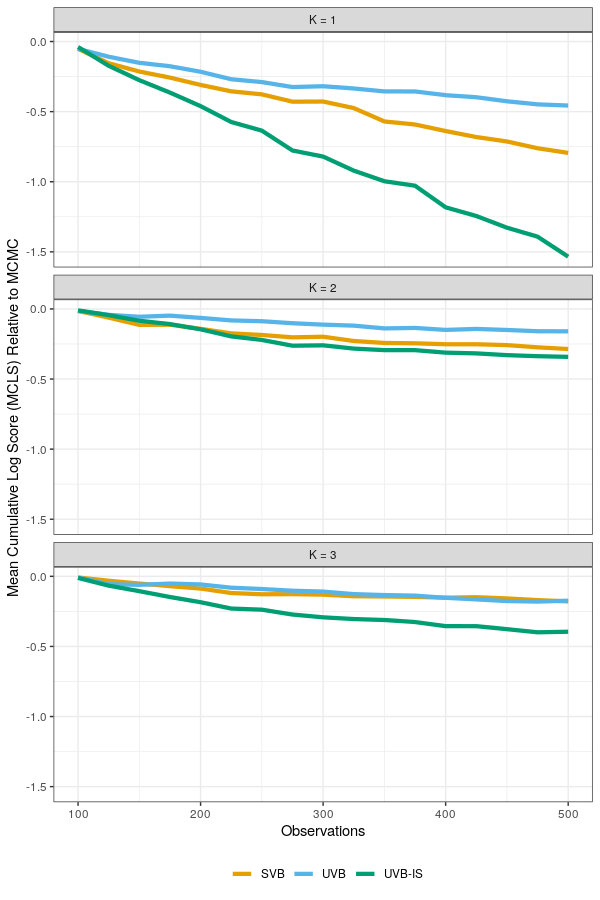}}
\hspace{1cm}
\subfloat[][Computational efficiency (RCMR)]{\includegraphics[width=7.5cm, height = 12cm]{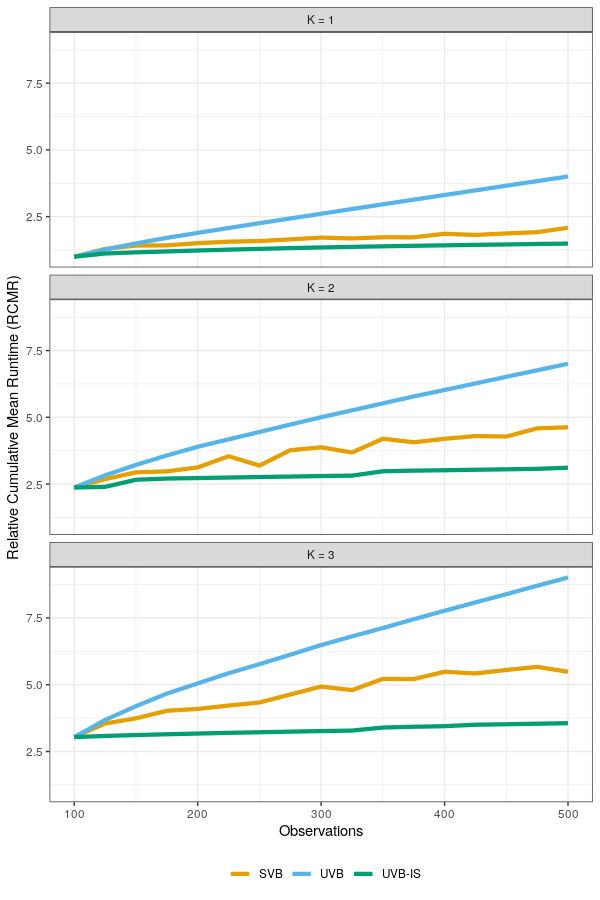} }
\caption{AR3 Simulation (a): Forecast accuracy, indicated by one-step-ahead mean cumulative predictive log scores (MCLS), corresponding to incremental updates under competing methods (SVB, UVB and UVB-IS) relative to MCMC. Higher values of MCLS indicate better forecast accuracy. (b): Computational efficiency, indicated by relative cumulative mean runtime (RCMR), again corresponding to incremental updates under competing method (SVB, UVB and UVB-IS), each reported relative to the mean runtime for SVB when $K=1$ and $T_n=100$.} 
\label{fig:UVBAR3Timing}%
\end{figure}

The forecast accuracy associated with the resulting approximate predictive densities is measured using the cumulative predictive log score $(CLS)$, given by
\begin{equation}
\label{UVB:TSlogscore}
CLS_n = \sum_{j = 1}^n\log(\widehat{p}(y^{(obs)}_{T_j + 1} | \bm y_{1:T_j})),
\end{equation}
for $n=1,2,...,21$, where $y^{(obs)}_{T_n + 1}$ denotes the realised (observed) value of $y_{T_n + 1}$. In particular, we compare the mean CLS (MCLS) over the $R$ replications, for each approximation method and each given value of $K$, at consecutive update times $T_n$ with $n=1,2,..,21.$ The results are displayed in Figure \ref{fig:UVBAR3Timing}, where each row indicates a different (known) value of $K$. Panel (a), on the left-hand side, the MCLS value is displayed relative to the MCLS value for MCMC inference, against the corresponding incremental values of $T_n+1 = 101, 126, \ldots, 501$. As greater values of MLS indicate better forecast accuracy, it is not surprising to find that each of the approximate VB method produces a lower MLS relative to exact (MCMC) inference. Amongst the approximate methods, UVB performs the best, in terms of MLS, followed by SVB and UVB-IS at $K=1$, though all of the variational methods improve as $K$ increases. 
\\

Panel (b) of Figure \ref{fig:UVBAR3Timing} displays the relative cumulative mean runtime (RCMR) for each VB method using data up to $T_n$, for $T_n = 100,125,150,\ldots 500$, each calculated by dividing the CMR by the mean run time of the SVB algorithm fitting a single mixture (i.e. $K=1$) at $T_n=100$. Note that the SVB approximation at each update time $T_n$ requires an application of the SGA algorithm using all data from time $T=1$ up to time $T_n$, while each of the updating methods are comprised of an SVB approximation at $T_1$, followed by $n-1$ progressive updates each using only the new data since the last update period. As can be seen in the top row of Panel (b), the RCMRs are all identical and equal to one at the first update time $T_n=100$ and all increase with consecutive updates. While all three methods show an increase in RCMR with each update, the UVB method appears least efficient. 
\\

In this setting the amount of data in each update is relatively small, and UVB increases the runtime compared to SVB. This is due to the computational overhead of $n$ SGA applications not being offset by a reduction in the number of log-likelihood calculations. In contrast, UVB-IS achieves sizeable computational gains despite showing minimal loss in the corresponding MLS for $K > 1$. 
\\

\begin{figure}[h]
\centering
\includegraphics[width = 0.95\textwidth]{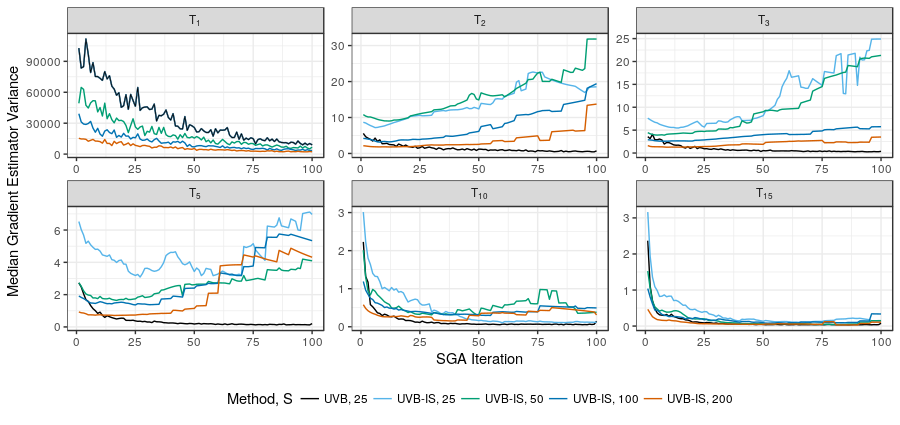}
\caption{Median gradient estimator variance for the first 100 SGA iterations, with colour indicated by the acronym (either UVB or UVB-IS) followed by the value of $S$. Both UVB and UVB-IS algorithms have arbitary starting values at $T_1$, denoted as $\bm \lambda_1^{(1)}$, where the estimated gradient exhibits high variance. Note the difference in the y-axis scales at subsequent update times, where the starting value at time $T_n$ is set to the previous optimal value, i.e. $\bm \lambda_n^{(1)} = \bm \lambda_{n-1}^*$, noting several time periods have associated $\bm \lambda^*_n$ values, but are omitted in the figure. Consequently, we the variance of the UVB gradient estimator is reduced relative to SVB, though the UVB-IS variance increases slightly for small $n$ and large iteration index $m$.}
\label{fig:UVBvariance}
\end{figure}

To illustrate the reduced variability in subsequent UVB gradient estimators, Figure \ref{fig:UVBvariance} displays the median variance of the gradient estimator for the posterior mean parameter of $\mu$, with UVB at $S = 25$, and UVB-IS at each of $S = 25, 50, 100$ and $200$, for each of six selected update periods. The algorithms at $T_1$ are applications of SVB with arbitary starting values for $\bm \lambda_1^{(1)}$. This causes extreme, but declining, variance until convergence is reached. This pattern is typical for SVB inference at all time periods, parameters, and values of $K$. In subsequent time periods each updating method sets the starting value at $\bm \lambda_n^{(1)} = \bm \lambda_{n-1}^*$, noting that, for example, $\bm \lambda_5^{(1)}$ is equal to the omitted $\bm \lambda_4^*$. The estimated variance is subsequently orders of magnitude smaller than SVB. For small values of $n$ the distributions $q_{\bm \lambda_n^*}(\bm \theta | \bm y_{1:T_n})$ and $q_{\bm \lambda_{n+1}^{(m)}}(\bm \theta | \bm y_{1:T_n})$ may differ as $m$ increases, causing a reduction in the effective sample size associated with the gradient estimator, and an increase in the UVB-IS estimator variance. This effect is visible at times $T_2, T_3$, and $T_5$, though the UVB-IS estimator variance is low relative to SVB despite this inefficiency.
\\

\subsection{Mixture Model Clustering}

\label{subsec:UVBMMC}

In the second simulated example we consider the case where repeated measurements are simulated on $N=100$ cross-sectional units at each of $T=100$ times. The measurements for a given unit follows one of two possible DGPs, with the objective being to cluster the units into the correct groups, according to the underlying DGP, with additional observations of each cross-sectional unit accumulating in an online fashion as time increases. Each of these scenarios was then replicated $R=500$ times.
\\

For each independent replication, we generate data $y_{i,t}$ as the measurement of unit $i$ at time $t$, for $i=1,2,...,N$ and $t=1,2,...,T$ as follows. We first define the cluster indicator for unit $i$ as $k_i$, and generate these for a given probability $0<\pi<1$ according to
\begin{equation}
\label{UVB:MMCkPrior}
k_i | \pi \stackrel{i.i.d.}{\sim} Bernoulli(\pi),
\end{equation}
where $i.i.d$ abbreviates \textit{independent and identically distributed.} Then, conditional on $k_i$ we let
\begin{equation}
\label{UVB:MMCmixNormalDGP2}
y_{i, t} | (k_i = j), \mu_j, \sigma^2_j \stackrel{ind}{\sim}  N(\mu_j, \sigma^2_{j}),
\end{equation}
for $j=0,1$, with $ind$ short for \textit{independent}. For this exercise, we set $\pi=0.5$, with the replicated values of $\mu_0$ and $\mu_1$ independently drawn from an $N(0, 0.25)$ distribution, while $\sigma_0^2$ and $\sigma_1^2$ are independently drawn from a uniform distribution over the interval $(1,2)$.
\\

Having simulated the data, the actual values $k_i$ are retained for each replication. We then use the UVB algorithm of the described model with the simulated data, as if all $N$ units were being observed online at increasing times $T_n=10, 20, 30, ..., 100$. The aim of the exercise is to cluster the units into two groups aligning with the true, but `unobserved' value of $k_i$.
\\

The Bayesian updating analysis proceeds as follows. Denoting the collective parameter vector as $\bm \theta = \{\log(\sigma^2_0), \log(\sigma^2_1), \mu_0, \mu_1 \}$, the joint prior for $\bm \theta$ and $\pi$ used at $T_1$ is given by independent components
\begin{align}
\bm \theta &\sim N(\boldsymbol{0}_4, 10 \mathbb{I}_4), \mbox{ and} \\
\pi &\sim Beta(\alpha, \beta). \label{UVB:MMCpiPriorMix}
\end{align}

Note that the model for $\pi$ in (\ref{UVB:MMCkPrior}) and the prior in (\ref{UVB:MMCpiPriorMix}) imply that the $k_i$ are independent \textit{a priori}, with marginal probabilities given by
\begin{equation}
\label{UVB:MMCkMarginalMix}
\Pr(k_i = j) = \frac{\mathcal{B}(j + \alpha, \beta - j + 1)}{\mathcal{B}(\alpha, \beta)},
\end{equation}
for $j=0,1$, where $\mathcal{B}(\cdot, \cdot)$ denotes the Beta function. Hence we have marginalised out the `unknown' value of $\pi$, and can now proceed to updating the prior in (\ref{UVB:MMCkMarginalMix}), for each $i=1,2,...,N$, on the basis of information at times $T_n=10,20,...,100.$
\\

Denoting $\bm y_{i, 1:T_n} = \{y_{i, t} | t = 1, \ldots, T_n\}$ and $\bm y_{1:N, 1:T_n} = \{\bm y_{i, 1:T_n} ; i = 1, \ldots N \}$,
the initial \textit{augmented} posterior distribution is given by
\begin{equation}
\label{UVB:MMCAugmented}
p(\bm \theta, \bm k_{1:N} | \bm y_{1:N, 1:T_1}) \propto p(\bm \theta) \prod_{i=1}^N p(\bm y_{i, 1:T_1} | \bm \theta, k_i = j) Pr(k_i = j),
\end{equation}
with each likelihood $p(\bm y_{i, 1:T_1} | \bm \theta, k_i = j)$ given by the product of densities associated with (\ref{UVB:MMCmixNormalDGP2}) and the value of $j$. 
\\

Due to the conditional independence of the components of $\bm \theta$ and the cluster indicators, subsequent posteriors at times $T_{n+1}$ are approximated by
\begin{equation}
\label{UVB:MMCupdateAug}
\widehat{p}(\bm \theta, \bm k_{1:N} | \bm y_{1:N, 1:T_{n+1}}) \propto  \prod_{i=1}^N p(\bm y_{i, T_n+1:T_{n+1}} | \bm \theta, k_i = j) \widehat{\Pr}(k_i = j | \bm y_{1:N, 1:T_{n}})p(\bm \theta |\bm  y_{1:N, 1:T_{n}}),
\end{equation}
where the latent class probabilities, $\Pr(k_i = j | \bm y_{1:N, 1:T_{n}})$, are estimated before updating with
\begin{equation} \label{UVB:mixk}
\widehat{\Pr}(k_i = j | \bm y_{1:N, 1:T_n}) \propto \frac{1}{M}\sum_{l=1}^M p(\bm y_{1:N, 1:T_n} | \bm \theta^{(l)}, k_i = j)\Pr(k_i = j),
\end{equation}
with $\bm \theta^{(l)} \sim p(\bm \theta | \bm y_{1:N, 1:T_n})$ for $l = 1, 2, \ldots, M$. 
\\

As in Section \ref{subsec:UVBTS}, the UVB and UVB-IS algorithms are compared to standard SVB and MCMC. Each of these approaches utilises an approximation to the augmented posterior of the form
\begin{equation}
\label{UVB:ApproxAugmented}
q_{\bm \lambda_{n+1}}(\bm \theta, \bm k_{1:N} | \bm y_{1:N, 1:T_{n+1}}) = q_{\bm \lambda_{n+1}}(\bm \theta | \bm y_{1:N, 1:T_{n+1}}) \prod_{i=1}^N \widehat{\Pr}(k_i = j | \bm y_{1:N, 1:T_n}),
\end{equation}
where $q_{\bm \lambda_{n+1}}(\bm \theta | \bm y_{1:N, 1:T_{n+1}})$ is a $K = 1, 2$, or $3$ component mixture of multivariate normal distributions and the $\bm \theta^{(l)}$ samples used to estimate (\ref{UVB:mixk}) are simulated from the previous approximation $q_{\bm \lambda_{n}}(\bm \theta | \bm y_{1:N, 1:T_{n}})$.
\\

The form of the approximation used in (\ref{UVB:ApproxAugmented}) is chosen due to the fact that the gradient of the augmented divergence, $KL[q_{\bm \lambda_{n}}(\bm \theta, \bm k_{1:N} | \bm y_{1:N, 1:T_{n}})\hspace{.1cm}||\hspace{.1cm}\widehat{p}(\bm \theta, \bm k_{1:N} | \bm y_{1:N, 1:T_{n}})]$ is equivalent to the gradient of the marginal divergence, $KL[q_{\bm \lambda_{n}}(\bm \theta | \bm y_{1:N, 1:T_{n}})\hspace{.1cm}||\hspace{.1cm}\widehat{p}(\bm \theta | \bm y_{1:N, 1:T_n})]$, and hence the same approximation can be found by instead targeting the marginal posterior distribution,
\begin{equation}
\label{UVB:MMCMarginal}
p(\bm \theta | \bm y_{1:N, 1:T_1}) \propto p(\bm \theta) \prod_{i=1}^N \left( \sum_{j=0}^1 p(\bm y_{i, 1:T_1} | \bm \theta, k_i = j) \Pr(k_i = j) \right),
\end{equation}
or its updated form
\begin{equation}
\label{UVB:MMCupdateMarginal}
\widehat{p}(\bm \theta | \bm y_{1:N, 1:T_{n+1}}) \propto  \prod_{i=1}^N \left( \sum_{j=0}^1 p(\bm y_{i, T_n+1:T_{n+1}} | \bm \theta, k_i = j) \widehat{\Pr}(k_i = j | \bm y_{1:N, 1:T_{n}})p(\bm \theta |\bm  y_{1:N, 1:T_{n}})\right).
\end{equation}

At each update we estimate class labels for $k_i$ according to
\begin{equation}
\widehat{k}_{i,n} = \arg \underset{j}{\max}\mbox{ } \widehat{\Pr}(k_i = j | \bm y_{1:N, 1:T_n}),
\end{equation}
and assign a classification accuracy (CA) score at $T_n$, given by
\begin{equation}
\label{MMCscore}
CA_n = \max \left(\frac{1}{N}\sum_{i=1}^N I(\widehat{k}_{i, n} = k_i), \hspace{1.5mm}\frac{1}{N}\sum_{i=1}^N I(\widehat{k}_{i, n} \neq k_i)\right),
\end{equation}
the proportion of successful classifications up to label switching.
\\
 
SVB and UVB gradients are estimated from $S = 25$ samples of $\bm \theta$ per iteration, while UVB-IS sets $S = 100$.
\\

\begin{figure}[h]
\subfloat[][Clustering Accuracy (MCA)]{\includegraphics[width=7.5cm, height = 12cm]{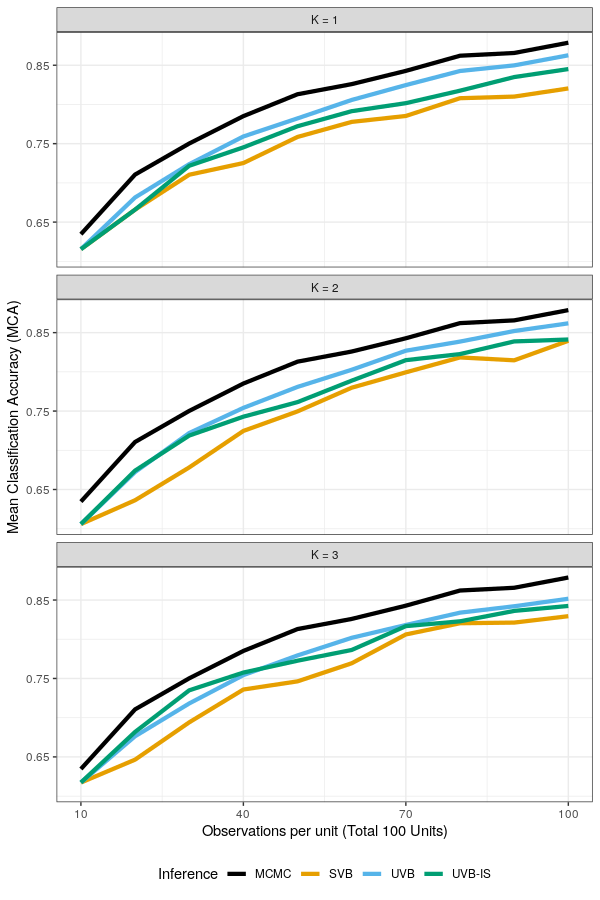}}
\hspace{1cm}
\subfloat[][Computational Efficiency (RCMR)]{\includegraphics[width=7.5cm, height = 12cm]{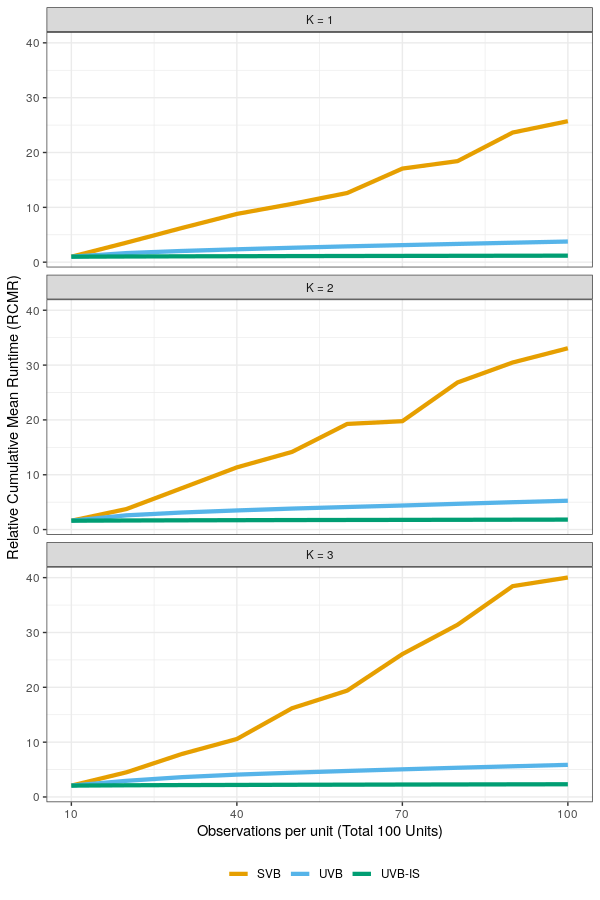}}
\caption{(a): Mean Classifcation Accuracy (MCA) for each inference method, higher is better. (b): Average SVB runtime for one approximation at time $T_n$ and average cumulative UVB and UVB-IS runtimes fit to $T_1$, then updated $n-1$ times to $T_n$. Runtimes are relative to the time required for the $T_1$ fit. UVB and UVB-IS perform better than SVB and are also much faster, as computation of the data likelihood is a large part of the gradient calculation in this scenario.}
	\label{fig:UVBMMCResults}%
\end{figure}

The results for this problem are displayed in Figure \ref{fig:UVBMMCResults}, where each row corresponds to a different value of $K$. Panel (a) displays the mean classification accuracy (MCA), corresponding to updates at times $T_n=10, 20, ..., 100$ and across $R$ replications. As in the previous study, each variational approximation reduces accuracy relative to exact inference. In this example both UVB and UVB-IS are more accurate than SVB, with little change apparent in any variational approach between different values of $K$.
\\

As in the previous section, Panel (b) of Figure \ref{fig:UVBMMCResults} displays the relative cumulative mean runtime (RCMR) for each VB method using data up to $T_n$, for $T_n = 10,20,\ldots, 100$, calculated relative to the mean run time of the SVB algorithm fitting a single mixture at the initial update time, when $T_n=10$. As the problem features a large number of cross-sectional units, the computational cost of calculating the log-likelihood dominates the gradient estimation. Processing smaller amounts of data, and having a reduced gradient variance, considerably increases the computational efficiency of UVB and UVB-IS relative to SVB while increasing accuracy. Despite the updating methods consisting of 10 SGA applications while SVB uses only one, UVB and UVB-IS require, on average, $14.7\%$, and $4.6\%$ of the computational time of SVB, respectively, in the top right panel when $K = 1$ at time $T_{10} = 100$.\\

\section{Eight Schools Example}\label{sec:eightschools}

In this section the co-called `Eight Schools' problem described in \citet{Gelman2014} is considered. This problem analyses the effectiveness of a short term coaching program, implemented independently by each of eight studied schools, for the SAT-V test\footnote{The SAT-V is a standardised aptitude test commonly taken by high school students in the USA.}. For students $i = 1, 2, \ldots, N_j$ in each school $j = 1, 2, \ldots, 8$, consider the linear regression
\begin{equation}
SAT\mbox{-}V_{i, j} = \beta_{0, j} + \beta_{1, j} Coach_{i, j} + \beta_{2, j} PSAT\mbox{-}V_{i, j} + \beta_{3, j} PSAT\mbox{-}M_{i, j} + \epsilon_{i, j}
\end{equation}
where $Coach_{i, j}$ is a dummy variable indicating a student's inclusion (or not) in a coaching program run by their school, alongside control variables $PSAT\mbox{-}V_{i, j}$ and $PSAT\mbox{-}M_{i, j}$, corresponding to each student's scores in the verbal and mathematical preliminary SAT, respectively.
\\

Following \citet{Gelman2014}, the estimated school-level coaching coefficients that correspond to the ordinary least squares estimators are taken as the observations, $y_j = \widehat{\beta}_{1, j},$ for $j = 1, 2, \ldots 8$, and have approximate sampling distributions given by 

\begin{equation}
\label{schools:CLT}
y_j |\theta_j, \sigma^2_j \sim \mathcal{N}(\theta_j, \sigma^2_j),
\end{equation}
where $\theta_j$ is the latent `true' effectiveness of school $j$'s coaching program. The standard deviation of the sampling distribution, $\sigma_j$, is assumed to be known and is held fixed at the standard error estimated by the relevant regression, with each having taken account of the individual school sample size $N_j$. 
\\

Again following \citet{Gelman2014}, we apply a hierarchical prior to the population mean values in (\ref{schools:CLT}), assuming that the $\theta_j$ themselves are random and $iid$ from a Student-$t$ distribution,
\begin{equation}
\label{schools:hier}
\frac{\theta_j - \mu}{\tau} \sim t(\nu)
\end{equation}

where $\nu$ is the degrees of freedom, fixed at $\nu = 4$. The hierarchical model also employs the uninformative hyper-prior
\begin{equation}
\label{schools:hyperprior}
p(\mu, \tau) \propto 1,
\end{equation}
over positive values of $\tau$, and both positive and negative values of $\mu$. 
\\

Collecting the unknown school means together and denoting by $\bm \theta_{1:8} = \{\theta_1, \theta_2, \ldots, \theta_8\}$, the posterior distribution of all unknowns and based on the observed values from all schools is then given by
\begin{equation}
\label{schools:posterior}
p(\bm \theta_{1:8}, \mu, \tau | \bm y_{1:8}) \propto \prod_{j=1}^8 p(y_j | \theta_j, \sigma^2_j) p(\theta_j | \tau, \mu).
\end{equation}

It is feasible to obtain this posterior exactly, via MCMC, for example using the algorithm provided in the statistical modelling platform \textit{Stan} \citep{RStanGettingStarted}.
\\

Our aim here is to demonstrate the application of UVB and UVB-IS to this hierarchical model, where with each update we sequentially `observe' an additional school, as indicated by the inclusion of an additional observation $y_j$. Each variational algorithm approximates the progressive posterior by the multivariate normal distribution $q_{\bm \lambda_n}(\bm \theta_{1:n}, \mu, \tau | \bm y_{1:n})$, for $n=1,2,...,8$. The initial distribution approximation at $T_1$ for UVB and UVB-IS is given by the multivariate normal distribution $q_{\bm \lambda_1^*}(\theta_1, \mu, \tau | y_1) $ where
\begin{equation}
\bm \lambda_1^* = \arg \underset{\bm \lambda_1}{\min}\hspace{1mm} KL[q_{\bm \lambda_{1}}(\theta_1, \mu, \tau |  y_1)  \hspace{.1cm}||\hspace{.1cm}p(\theta_1, \mu, \tau |  y_1)].
\end{equation}
Updates at further `times' $T_{n+1}=n+1,$ for $n = 1, 2, \ldots, 7$, involves sequentially adding schools to the model targetting the pseudo-posterior distribution, given by the decomposition
\begin{equation}
\label{schools:update}
\widetilde{p}(\bm \theta_{1:n+1}, \mu, \tau | \bm y_{1:n+1}) \propto p(y_{n+1}| \theta_{n+1})p(\theta_{n+1} | \mu, \tau)q_{\bm \lambda_n^*}(\bm \theta_{1:n}, \mu, \tau | \bm y_{1:n}).
\end{equation}

Either UVB or UVB-IS then may be used to obtain the updated approximate posterior, given by $q_{\bm \lambda_{n+1}^*}(\bm \theta_{1:n+1}, \mu, \tau | \bm y_{1:n+1})$, with 
\begin{equation}
\bm \lambda_{n+1}^* = \arg \underset{\bm \lambda_{n+1}}{\min}\hspace{1mm} KL[q_{\bm \lambda_{n+1}} (\bm \theta_{1:n+1}, \mu, \tau | \bm  y_{1:n+1}) \hspace{.1cm}||\hspace{.1cm}\widetilde{p}(\bm \theta_{1:n+1}, \mu, \tau | \bm  y_{1:n+1})],
\end{equation}
for $n=1,2,...,y.$ As each update adds a new variable $\theta_{n+1}$ to the model, the optimal vector $\bm \lambda_{n+1}^*$ updates the auxiliary parameters associated with the pseudo-posterior distribution for $\theta_{n+1}$ together with the previously included variables $\mu, \tau$, and $\bm \theta_{1:n}$. We note that our implementation of UVB-IS here employs a hybrid strategy utilising importance sampled gradients (\ref{UVBIS:scoreEstIS}) for simulations of $\mu, \tau$, and $\bm \theta_{1:n}$ from the previous $q_{\bm \lambda_{n}^*}(\bm \theta_{1:n}, \mu, \tau | \bm y_{1:n+1})$, and score-based gradients for $\bm \theta_{n+1}$, as per (\ref{scoreDeriv}). The score-based gradients use samples generated from $\bm \theta_{n+1} \sim q_{\bm \lambda_{n+1}}(\theta_{n+1} |\bm  y_{1:n+1}, \mu, \tau, \bm \theta_{1:n})$, which is available as this variational approximation was chosen to be a multivariate normal distribution.
\\

We compare approximations that result from using UVB and UVB-IS, relative to the sequential implementation of SVB, as each new school is added. As the ordering of the inclusion of schools is arbitrary in this example, we report results that are averaged over a randomly selected $100$ of the $8! = 40,320$ possible permutations of school sequences. For each ordering, the KL divergence $KL[q_{\bm\lambda}(\bm {\theta}_{1:n+1}, \mu, \tau |\bm y_{1:n+1} )\hspace{.1cm}||\hspace{.1cm}p(\bm {\theta}_{1:n+1}, \mu, \tau |\bm y_{1:n+1})]$ is calculated and tallied, using the exact posterior $p((bm {\theta}_{1:n+1}, \mu, \tau |\bm y_{1:n+1})$ in (\ref{schools:posterior}), each calculated using 10,000 MCMC sample draws retained following a burn-in period of 10,000 iterations.
\\

These results are shown in Table \ref{table:schools}, where the row labelled UVB contains the average difference between the KL divergence calculated from the UVB approximation and the KL divergence calculated from the SVB approximation. These values are calculated independently for each margin of the distribution, and for the entire joint distribution. Each KL divergence was obtained using simulated draws from each relevant distribution. Marginal densities were estimated using the density function in R \citep{R}, with the corresponding joint density estimated using an additional vine copula \citep{Dissmann2013}. The row labelled UVB-IS contains the average difference between the corresponding UVB-IS and SVB KL divergences.

\begin{table}[htbp]
\centering
\begin{tabular}{|l|l|l|l|l|l|l|l|l|l|l|l|}
\hline
 & $\tau$ & $\mu$ & $\theta_1$ & $\theta_2$ & $\theta_3$ & $\theta_4$ & $\theta_5$ & $\theta_6$ & $\theta_7$ & $\theta_8$ & Joint \\
 \hline
UVB & 1.04 & \textbf{0.11} & \textbf{0.26} & \textbf{0.05} & \textbf{0.16} & \textbf{0.08} & \textbf{0.16} & \textbf{0.12} & \textbf{0.16} & \textbf{0.14} & \textbf{2.67}\\
\hline
UVB-IS & \textbf{0.16} & 0.64 & 0.32 & 0.34 & 0.38 & 0.22 & 0.17 & 0.27 & 0.36 & 0.42 & 5.00\\
\hline
\end{tabular}
\caption{Results of the Schools example. Average KL divergence measures: $KL_{UVB} - KL_{SVB}$ (row 2) and $KL_{UVB-IS} - KL_{SVB}$ (row 3), with each KL divergence calculated from the VB approximation to MCMC samples. Lower is better.}
\label{table:schools}
\end{table}

\section{Lane Position Example}\label{sec:lanepos}

\begin{figure}[h]
\centering
\includegraphics[width=0.95\textwidth]{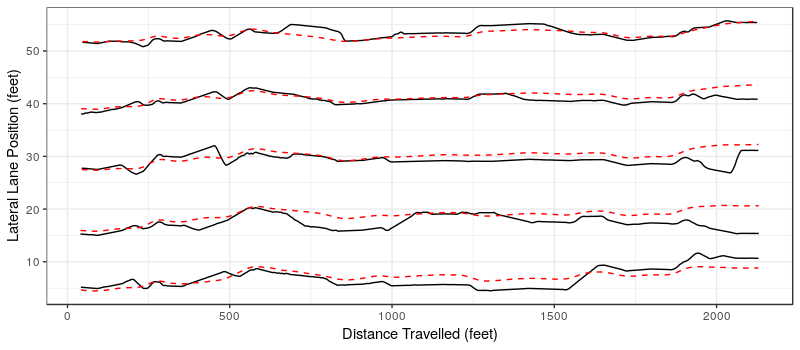}
\caption{The path of five selected vehicles from the NGSIM dataset, travelling from left to right, with each black line representing a unique vehicle, with estimated lane centre lines in red. This section of US Route 101 is comprised of five main lanes, with a sixth entry/exit lane not shown.}
\label{fig:carsMidlane}
\end{figure}

Vehicle drivers may exhibit a tendency to move laterally (i.e. side-to-side) within their designated lane on a highway. Figure~\ref{fig:carsMidlane} displays this notion, by plotting the trajactory of five drivers as they travel along a section of the US Route 101 Highway, as taken from the Next Generation Simulation (NGSIM, \cite{NGSIM2017}) dataset. In this figure, the vehicles - indicated in black - are travelling towards the right, with each (estimated) lane centre line given by the red dashed line. Drivers likely adapt their position in real time, in at least partial response to the perceived position of vehicles that are travelling nearby.
\\

The aim of this section is to apply the UVB methodology to analyse a model of the lateral position of vehicles. The model incorporates driver heterogeneity, while the analysis itself produces sequential, per-vehicle distributional forecasts of a large number of future car positions. The methodology suggests that a smart vehicle (i.e. one without a human driver) may be able to repeatedly `observe' neighbouring vehicle positions, predict their positions in real time as they travel along the road, and appropriately respond to those forecasts.
\\

To set up the scenario, we randomly select, from the NGSIM dataset, trajectories associated with $N=500$ vehicles that do not change lane. We note that the NGSIM dataset is the result of a project conducted by the US Federal Highway Administration (FHWA), and includes data recorded from $6101$ vehicles traveling along a 2235 foot long section of the US 101 freeway in Los Angeles, California from 7:50 am to 8:35 am on June 15th, 2005. Though initially collected by static cameras, the data was then processed by Cambridge Systematics Inc. to produce coordinates of the centre of the front of each vehicle at 100 millisecond intervals.
\\

\subsection{A Hierarchical Model}
In developing a model for the position of cars we consider a number of issues. First, each vehicle/driver is likely to have its own idiosyncratic behaviour, captured by its own parameter values. Let $y_{i,t}$ denote the lateral deviation from the lane centre of vehicle $i$ at time $t$, with details on calculating the lateral deviation provided in Appendix \ref{AppA}. For $i=1,\ldots,N$ and $t=1,\ldots,T$, we assume
\begin{equation}
y_{i, t}  \mid \mu_i, \sigma^2_i \overset{ind}{\sim} \mathcal{N}\left(\mu_i, \sigma_i^2\right),\label{eq:DGP}
\end{equation}
where $\mu_i$ and $\sigma_i^2$ are parameters specific to vehicle $i$. For simplicity, we collect the individual vehicle-specific parameters into a single vector, $\boldsymbol{\theta}_i$, by defining $\boldsymbol{\theta}_i = (\mu_i,\log{(\sigma^2_i)})$, for $i=1,2,..,N$. We note that alternative parametric models could be used here, including a time series model for vehicle $i$, with little loss in generality.\\

Multiple cars may display similar behaviour, a phenomenon that can be modeled by allowing different cross sectional units to share parameters. This structure, whereby cross sectional units belong to mixture components, leads to predictions that `borrow strength' from the full sample of vehicles. To make this idea explicit, let $k_i$ denote an indicator variable such that vehicle $i$ belongs to mixture component $j$ if $k_i=j$. All vehicles within the same mixture component share parameters, that is $\theta_i=\theta^*_j$, for all $i$ such that $k_i=j$. Note that the star superscript and $j$ subscript are generally used to index the mixture component that the parameters belong to, while the subscript $i$ is generally used to index the cross-sectional unit, i.e. vehicle.\\

Since the number of components are unknown and since there is a possibility that a new vehicle will be observed with behaviour that cannot be well described by any of the prevailing parameters, we consider an infinite mixture model. In particular, we use an infinite mixture model induced by a Dirichlet Process (DP) Prior for the distribution of the parameters. The DP prior is given by
\begin{equation}
G \sim DP(\alpha, G_0)\label{eq:DP}\,,
\end{equation}
where $G_0$ is the DP base distribution, assumed here to be $N(\boldsymbol{0}_2, 10 \mathbb{I}_2)$, and the DP concentration parameter $\alpha$ is fixed here and equal to one. The prior for the collection of $\boldsymbol{\theta}_i$ values represent a draw from the DP, with
\begin{equation}
\bm \theta_i | G \overset{iid}{\sim} G, \mbox{ for } i = 1, 2, \ldots, N \label{eq:DPPrior}\,.
\end{equation}
Combining (\ref{eq:DGP}), (\ref{eq:DP}) and (\ref{eq:DPPrior}) leads to the hierarchy
\begin{align}
G &\sim DP(\alpha, G_0)\nonumber\\
\bm \theta_i | G &\overset{iid}{\sim} G, \mbox{ for } i = 1, 2, \ldots, N\nonumber\\ y_{i, t} \mid \bm \theta_i &\overset{ind}{\sim} \mathcal{N}\left(\mu_i, \sigma_i^2\right), \, \mbox{ for } i = 1, 2, \ldots, N \mbox{ and } t=1,2, \ldots, T.\label{eq:hierarchy}
\end{align}

We note that the DP prior induces clustering on the observation sequences, as described by the Chinese Restaurant Process (CRP, \citealp{Aldous1985}) representation. The CRP provides a mechanism for drawing from the prior of $\bm \theta_1,\ldots, \bm \theta_n$, marginal of the random $G$, via the introduction of discrete variables that act as component indicators. Define $s_i$ as the number of unique values in $k_1, k_2, \ldots, k_i$, and let $n_{ij} = \sum_{m=1}^i I(k_m = j)$. Then, the indicator variables can be simulated from $p(k_i = j | \alpha, \bm k_{1:i-1} )$ where
\begin{align}
\label{DPMprobs}
p(k_1 = 1 \mid \alpha, G_0) &= 1, \\
p(k_i = j | \alpha, G_0,~ \bm k_{1:i-1}) &= \left\{ \begin{array}{cc} \frac{n_{i-1, j}}{\alpha + i - 1} & \hspace{6mm} \mbox{for } j = 1, 2, \ldots, s_{i-1} \\
\frac{\alpha}{\alpha + i - 1} & \mbox{for } j = s_{i-1}+1, \end{array} \label{DPMprobs2} \right.
\end{align}
for $i = 2, \ldots, N$. Note that although simulation of the indicators does not require knowledge of $G_0$, we include explicit conditioning on both $\alpha$ and $G_0$ in (\ref{DPMprobs}) and (\ref{DPMprobs2}) to emphasise the marginalisation over $G$. Under the CRP, unique values of $ \bm \theta_i$, denoted as $\bm \theta_j^*$, for $j=1,2,...,s_N$ are drawn from the base distribution $G_0$, and if we set $\bm \theta_i=\bm \theta^*_j$ for all $i$ such that $k_i=j$, then $(\bm \theta_1,\bm \theta_2,\ldots,\bm \theta_N)$ is a draw from the hierarchical setup in (\ref{eq:hierarchy}). Note that although the model is an infinite component mixture model, under the CRP the maximum number of unique clusters, $s_N$, can be no greater than the number of vehicles in the sample, $N$. For simplicity we retain the full vector $\bm \theta^*_{1:N}$, noting that some values $\bm \theta^*_{1:N}$ may not be associated with any vehicle.\\

The overall model may be seen as a Dirichlet Process Mixture (DPM) model for the lane deviations. Background material regarding Bayesian analysis of DPM models, including many references and detailed discussions relating to MCMC-based techniques for sampling from the relevant posterior, is given in \cite{Muller2015}. Online VB-based inference for DPMs has been established using a Mean Field approach - see, e.g., \cite{Hoffman2010}, \cite{Wang2011}, and \cite{Kabisa2016}). In contrast, our approach incorporates SVB, which allows for greater flexibility regarding the form of the approximating posterior distribution. Another important distinction between our analysis and this literature is that our setting involves multiple observations over time, on each cross-sectional unit. Rather than updating as new cross-sectional units are observed, we update parameters relating to the same cross-sectional units observed periodically over a period of time.
\\

\subsection{Implementation of SVB at time $T_1$}\label{sec:UVBatT1}
Before discussing how UVB is applied to this problem it is instructive to discuss how SVB is implemented for the DPM in (\ref{eq:hierarchy}) that targets the posterior conditional on all cross-sectional units $N$ over just the first time period from $t=1$ to $t=T_1$. For notational convenience, conditional dependence on $\alpha$ and $G_0$ is supressed in all notation for the remainer of this section. The objective is to minimise the KL divergence between a suitable variational approximation and a posterior that is augmented by indicator variables. To implement SVB, we must evaluate
\begin{equation}
p( \bm y_{1:N,1:T_1}, \bm \theta_{1:N}^*, \bm{k}_{1:N})=\left[\prod\limits_{i=1}^{N}\prod\limits_{t=1}^{T_1} p(y_{i,t}| \bm \theta_{1:N}^*, k_i)\right]\left[\prod\limits_{i=1}^{N}p(k_i|\bm k_{1:i-1})\right] p(\bm \theta^*_{1:N})\, 
\label{DPMtarget}
\end{equation}
for given values of $ \bm y_{1:N,1:T_1}, \bm \theta_{1:N}^*$, and $\bm{k}_{1:N},$. Each of the three main components on the right hand side of (\ref{DPMtarget}) can be computed from the hierarchical structure in~(\ref{eq:hierarchy}) and the CRP, as \citet{Sethuraman1994} shows that the unique values $\bm \theta^*_{1:N}$ are \textit{a priori} independent and identically distributed according to the base distibution $G_0$.
\\

A second required input into SVB is an approximate posterior density structure, given by $q_{\bm \lambda}$, and for this we propose
\begin{equation}
q_{\bm \lambda}(\bm \theta_{1:N}^*, \bm k_{1:N} | \bm y_{1:N, 1:T_1}) = \left[ \prod_{j=1}^N q_j(\theta_j^* | \bm y_{1:N, 1:T_1}) \right] \left[\prod_{i=1}^Np(k_i | \bm y_{1:N, 1:T_1}, \bm \theta^*_{1:N}, \bm k_{1:i-1}) \right]. \,
\label{DPMapprox}
\end{equation}
Each $q_j(.)$ on the right hand side is a bivariate normal distribution with unique means, variances and covariances for each $i=1,2,..,N$, leading to a total of $5N$ auxiliary parameters in the approximation. In the second product term on the right hand side of (\ref{DPMapprox}), the notation $p$ is used instead of $q$ since $p(k_i | \bm y_{1:N, 1:T_1}, \bm k_{1:i-1}, \bm \theta^*_{1:N})$ is known exactly and can be computed recursively using 
\begin{equation}
\label{DPMprobspost}
p(k_i = j | \bm y_{1:N, 1:T_1}, \bm k_{1:i-1}, \bm \theta^*_{1:N}) \propto p(k_i = j | \bm k_{1:i-1})p(\bm y_{i, 1:T_1} | \bm \theta^*_{1:N}, k_i),
\end{equation}
for $i=1,2,...,N$.
\\

The use of the so-called full conditional distribution for $\bm k_{1:N}$, given by the second product in (\ref{DPMapprox}), is a novel inclusion that enables our model to capture some of the dependence structure of the posterior. In contrast, a MFVB approximation would force posterior independence between each $k_i$ and every $\theta_j^*$, as in, for example, \cite{Wang2011}.
\\

Furthermore, in addition to minimising the KL divergence to the augmented posterior, our choice has the benefit of ensuring minimisation of the KL divergence to the corresponding marginal posterior. That is, the augmented gradients are given by
\begin{equation}
\frac{\partial KL[q_{\bm \lambda_{1}}(\bm \theta^*_{1:N}, \bm k_{1:N} | \bm y_{1:N, 1:T_1})\hspace{.1cm}||\hspace{.1cm}p((\bm \theta^*_{1:N}, \bm k_{1:N} | \bm y_{1:N, 1:T_1})]}{\partial \bm \lambda_{1}}
\end{equation}
and are equal to the marginal gradients 
\begin{equation}
\frac{\partial KL[q_{\bm \lambda_{1}}(\bm \theta^*_{1:N}| \bm y_{1:N, 1:T_1})\hspace{.1cm}||\hspace{.1cm}p(\bm \theta^*_{1:N} | \bm y_{1:N, 1:T_1})]}{\partial \bm \lambda_1},
\end{equation}
and so the optimisation procedure is equivelent to one where the indicator variables used to construct the DPM have been marginalised out.
\\

The proof of this result is shown in Appendix \ref{AppB}.

\subsection{Iterating UVB}\label{sec:IteratingUVB}

Using data up to time $T_1$, the first UVB posterior is obtained using SVB, as described in Section \ref{sec:UVBatT1}. For updating at time $T_{n+1}$, we construct a pseudo-posterior using information from the previous variational approximation $q_{\bm \lambda_n}(\bm \theta_{1:N}^*, \bm k_{1:N} | \bm y_{1:N,1:T_{n}})$ in two distinct ways. First, the base distribution in the DP as the prior distibution for $\bm \theta_{1:N}^*$ is updated to reflect the clustering present in the previously obtained posterior, and so is replaced with $q_{\bm \lambda_{n}}(\bm \theta^*_{1:N} |\bm y_{1:N,1:T_{n}})$. Second, retaining the form of the approximation in (\ref{DPMapprox}) for the update is complicated by the use of the full conditional distribution for $k_i$, given by
\begin{equation}
p(k_i = j | \bm y_{1:N, 1:T_{n+1}},\bm \theta^*_{1:N}, \bm k_{1:i-1}) \propto p(\bm y_{i, T_{n}+1:T_{n+1}} | \bm \theta^*_{1:N}, k_i)p(k_i = j | \bm y_{1:N, 1:T_{n}}, \bm \theta^*_{1:N}, \bm k_{1:i-1}),
\end{equation}
as all currently observed data up to time $T_{n+1}$ is required for each new $\bm \theta_{1:N}^*$ value simulated within the SGA algorithm. Instead our approach is to marginalise the variational distribution using
\begin{equation}
\label{DPMprobsmarginal}
q(k_i = j | \bm y_{1:N, 1:T_{n}}, \bm k_{1:i-1}) = \int_{\bm \theta_{1:N}^*} q_{\bm \lambda_{n}}(\bm \theta_{1:N}^*| \bm y_{1:N, 1:T_{n}})p(k_i = j | \bm y_{1:N, 1:T_{n}}, \bm \theta^*_{1:N}, \bm k_{1:i-1})d \bm \theta_{1:N}^*,
\end{equation}
before each update, estimating (\ref{DPMprobsmarginal}) from a sample average of $p(k_i = j | \bm y_{1:N, 1:T_{n}}, \bm \theta^*_{1:N}, \bm k_{1:i-1})$ using $M$ samples $\bm \theta_{1:N}^*$ and $\bm k_{1:i-1}$ simulated from the available approximate distribution. This requires use of all observed data at $T_n$, for each of the $M$ samples, but is independent of $\bm \theta_{1:N}^*$ and thus data up to $T_n$ is not required as new $\bm \theta_{1:N}^*$ values are simulated in the SGA algorithm. The component of the variational approximation for $k_i$ is then replaced by
\begin{equation}
\label{DPMprobspost2}
\widehat{p}(k_i = j| \bm y_{1:N, 1:T_{n+1}},\bm \theta^*_{1:N}, \bm k_{1:i-1}) \propto p(\bm y_{i, T_{n}+1:T_{n+1}} | \bm \theta_{1:N}^*, k_i = j)q(k_i = j | \bm y_{1:N, 1:T_{n+1}}, \bm k_{1:i-1}),
\end{equation}
which may be calculated using only the newly observed data $\bm y_{1:N, T_n+1:T_{n+1}}$ in the SGA algorithm. Note that the marginalisation step for all updates uses the exact full conditional distribution from the CRP representation, $p(k_i = j | \bm y_{1:N, 1:T_{n}},\bm \theta^*_{1:N}, \bm k_{1:i-1})$, rather than the marginalised form $\widehat{p}(k_i = j| \bm y_{1:N, 1:T_{n+1}}, \bm \theta^*_{1:N}, \bm k_{1:i-1})$ from the previous update.
\\

The targeted pseudo-posterior distribution for the update at $T_{n+1}$ is given by
\begin{equation}
\label{carUpdate2}
\widetilde{p}(\bm \theta^*_{1:N}, \bm k_{1:N} | \bm y_{1:N, 1:T_{n+1}}) \propto \prod_{i=1}^N \left[p(\bm y_{i, T_{n}+1:T_{n+1}} | \bm \theta_{1:N}^*, k_i) q(k_i | \bm y_{1:N, 1:T_{n}}, \bm k_{1:i-1})\right] q_{\bm \lambda_{n}^*}(\bm \theta_{1:N}^* | \bm y_{1:N, 1:T_{n}}),
\end{equation}
where the base distribution of the DP posterior in the DPM (and its corresponding CRP) is replaced with it's associated variational approximation at time $T_n$. The approximating distribution for the update at time $T_{n+1}$ is given by
\begin{equation}
q_{\bm \lambda_{n+1}}(\bm \theta_{1:N}^*, \bm k_{1:N} | \bm y_{1:N, 1:T_{n+1}}) = \prod_{j=1}^N q_{j,n+1}(\theta_j^* | \bm y_{1:N, 1:T_{n+1}})\prod_{i=1}^N \widehat{p}(k_i | \bm y_{1:N, 1:T_{n+1}}, \bm \theta^*_{1:N}, \bm k_{1:i-1}).
\label{DPMapproxUpdate}
\end{equation}

Given the pseudo-posterior (\ref{DPMtarget}), form of approximating distribution (\ref{DPMapproxUpdate}), and components of the time $T_n$ approximation: $q_{\bm \lambda_{n}}(\bm \theta_{1:N}^*| \bm y_{1:N, 1:T_{n}})$ and $q(k_i = j | \bm y_{1:N, 1:T_{n}}, \bm k_{1:i-1})$, the optimal parameter vector at time $T_{n+1}, \bm \lambda^*_{n+1}$, may be obtained via Algorithm \ref{alg:DPM}.

\vspace{2mm}
\begin{algorithm}[htbp]
\SetKwInOut{Input}{Input}
\Input{DP base distribution $G_0$ or updated approximating distribution at $T_n$.}
\KwResult{Approximating distribution at $T_{n+1}$.}
Calculate (\ref{DPMprobsmarginal}) for all $i$.\;
Observe $\bm y_{1:N, T_{n}+1:T_{n+1}}.$\;
Set $\mathcal{L}(q, \bm \lambda_{n+1}^{(0)}) = - \infty$.\;
Set initial values $\bm \lambda_{n+1}^{(1)}$.\;
Set $m = 1$.\;
\While{$|\mathcal{L}(q, \bm \lambda_{n+1}^{(m)}) - \mathcal{L}(q, \bm \lambda_{n+1}^{(m-1)})| < \epsilon$}{
	Simulate $\theta^{*(s)}_{1:N} \sim q_{\bm \lambda_{n+1}^{(m)}} (\bm \theta_{1:N}^* | \bm y_{1:N, 1:T_{n+1}})$ for $s = 1, 2, \ldots, S$.\;
	Simulate $\bm k_{1:N}^{(s)}$ with probabilities (\ref{DPMprobspost}) or (\ref{DPMprobspost2}).\;
	Evaluate $\widetilde{p}(\bm y_{1:N, 1:T_{n+1}}, \bm \theta^{*(s)}_{1:N}, \bm k_{1:i-1}^{(s)})$.\;
	Evaluate $q_{\bm \lambda_{n+1}}(\bm \theta_{1:N}^{*(s)}, \bm k_{1:N}^{(s)} | \bm y_{1:N, 1:T_{n+1}})$.\;
	Evaluate $\partial q_{\bm \lambda_{n+1}}(\bm \theta_{1:N}^{*(s)}, \bm k_{1:N}^{(s)} | \bm y_{1:N, 1:T_{n+1}}) / \partial \bm \lambda_{n+1}$.\;
	Update auxiliary parameter $\bm\lambda_{n+1}^{(m+1)} = \bm\lambda_{n+1}^{(m)} + \rho^{(m)} \widehat{\frac{\partial\mathcal{L}(q, {\bm\lambda_{n+1}})}{\partial {\bm\lambda_{n+1}}}} \bigg\rvert_{{\bm\lambda_{n+1}} = {\bm\lambda_{n+1}}^{(m)}}$.\;
	Calculate $\mathcal{L}(q, \bm \lambda_{n+1}^{(m+1)})$.\; 
	\vspace{0.2cm}
	Set $m = m+1$.\;
}
\caption{UVB for the DPM}
\label{alg:DPM}
\end{algorithm}
 
\subsection{Predicting Lane Positions}
\label{sec:UVBpredicting}
 
Given a posterior approximation $q_{\bm \lambda_{n}}(\bm \theta_{1:N}^*, \bm k_{1:N} | \bm y_{1:N, 1:T_{n}})$ we may obtain the approximate predictive distribution for vehicle $i$ at some future time $T_n+h$ as
\begin{equation}
\label{predDPM}
q(y_{i, T_n+h} | \bm y_{1:N, 1:T_n}) = \int p(y_{i, T_n+h} | \bm \theta_{1:N}^*, k_i) q_{\bm \lambda_{n}}^{*} (\bm \theta_{1:N}^*, \bm k_{1:N} | \bm y_{1:N, 1:T_{n}}) d \bm \theta_{1:N}^* d \bm k_{1:N}.
\end{equation}
After obtaining this distribution from samples $\{\bm \theta^*_{1:N}, \bm k_{1:N}\}^{(j)} \sim q_{\bm \lambda_{n}}^{*}(\bm \theta_{1:N}^*, \bm k_{1:N} | \bm y_{1:N, 1:T_{n}}),$ for $j = 1, 2, \ldots, M$, we calculate the predictive log score (LS),
\begin{equation}
LS_{i, n, h} = \log(q(y_{i, T_n+h}^{(obs)} | \bm y_{1:N, 1:T_n})),
\end{equation}
where $y_{i, T_n+h}^{(obs)}$ is the observed value of $y_{i, T_n+h}$. The performance of the UVB algorithm is evaluated by comparing its cumulative predictive log scores relative to those produced by competing methods.
\\

We also infer the DPM model via MFVB using the so-called `stick-breaking' representation of the Dirichlet Process, as in \cite{Wang2011}. This approach estimates the fully factorised posterior approximation, given by
\begin{equation}
q_{\bm \lambda}(\bm \theta_{1:N}^*, \bm k_{1:N} | \bm y_{1:N, 1:T_n}) = \prod_{j=1}^N q(\theta_j^*  | \bm y_{1:N, 1:T_n})q(k_j  | \bm y_{1:N, 1:T_n}).
\end{equation}
This may be used to build a predictive distribution in the same manner as (\ref{predDPM}). Details of the MFVB approximation are provided in Appendix \ref{AppC}.
\\

To illustrate the benefits of including posterior dependence in the approximation, we also introduce a parametric and \textit{independent} model, which retains a normal likelihood for each vehicle, i.e.
\begin{equation}
y_{i, t} \sim \mathcal{N}(\mu_i, \sigma^2_i)
\end{equation}
and assumes for each vehicle an independent uninformative prior, given by
\begin{equation}
p(\mu_i, \sigma^2_i) \propto \sigma^{-2}_i.
\end{equation}
For this model the predictive distribution for vehicle $i$ is analytically available as
\begin{equation}
\label{predT}
p(y_{i, T_n+h} | \bm y_{i, 1:T_n}) = \frac{\Gamma\left(\frac{\nu + 1}{2}\right)}{\Gamma\left(\frac{\nu}{2}\right)\sqrt{\frac{\pi \nu (T_n + 1) s^2_{i, n}}{T_n}}} \left(1 + \frac{T_n\left(y_{i, T_n+h} - \bar{y}_{i, n} \right)^2}{\nu (T_n+1)s^2_{i, n}} \right)^{\frac{-(\nu + 1)}{2}}
\end{equation}
a location-scale transform of the usual Student-$t$ distribution with $\nu = T_n-1$ degrees of freedom, where $\bar{y}_{i, n}$ and $s^2_{i, n}$ denote the sample mean and variance of $\bm y_{i, 1:T_n}$, respectively. Note that this model ignores any information from all other vehicles, and is similarly evaluated by the corresponding cumulative predictive log score.

\subsection{Analysis of the NGSIM Data}
We now discuss the empirical application results from the UVB algorithm for the DPM model described above for the NGSIM data. The posterior updates for both the cluster locations, $\bm \theta^*_{1:N}$, and the indicator variables, $\bm k_{1:N}$, occur at a sequence of pre-determined time periods, given by $T_1 = 50, T_2 = 75, T_3 = 100, T_4 = 125, T_5 = 150$, and $T_6 = 175$. \\

\begin{figure}[htbp]
	\centering
	\subfloat[][Weighted $\bm \theta^*$ marginal posterior approximation, based on UVB, at time $T_6$.]{\includegraphics[width=16cm, height = 6cm]{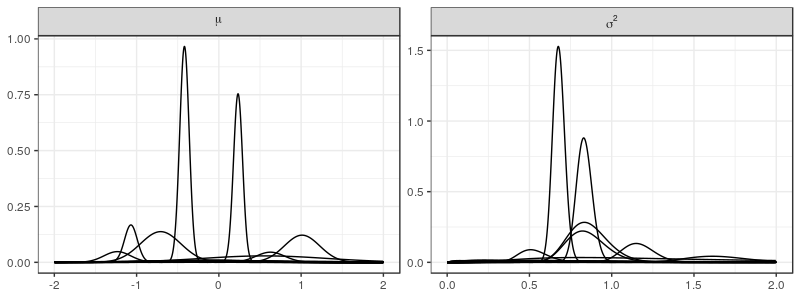}}
	\hspace{1cm}
	\subfloat[][UVB Predictive moments for high probability groups, at time $T_6$.]{\includegraphics[width=16cm, height = 6cm]{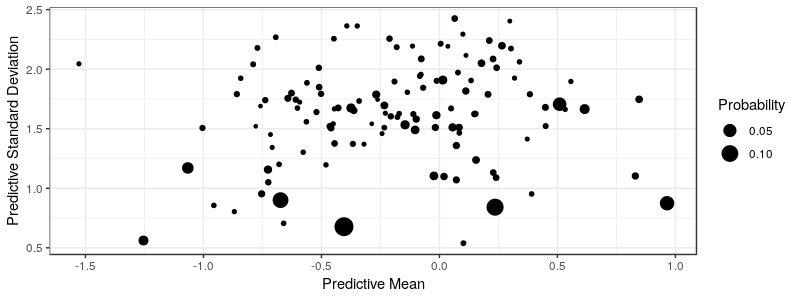}}
	\hspace{1cm}
	\subfloat[][Individual vehicle and average predictive densities from UVB at time $T_6$.]{\includegraphics[width=16cm, height = 6cm]{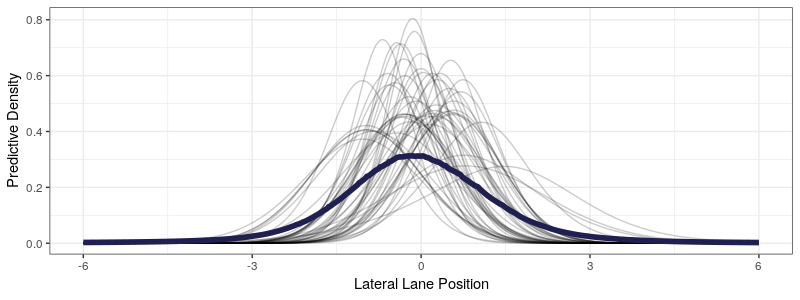}}
	\caption{(a): Posterior approximation for each $\bm \theta^*$, weighted by proportion of $\bm k_{1:N}$ draws. Two groups have high posterior precision with numerous groups showing more uncertainty. (b): Posterior predictive distribution means and standard deviations, sized according to the top 80\% of $\bm k_{1:N}$ draws. (c): Averaged predictive distribution for all groups in dark blue, with a random subset of fifty per vehicle distributions in grey.}%
	\label{fig:DPPresults}%
\end{figure}

Consider first the two graphs shown in the top panel (panel (a)) of Figure \ref{fig:DPPresults}. In each graph, the approximate marginal posterior distributions for each unique value $\mu^*_j$ (on the left) and $\sigma^{2, *}_j$ (on the right). Noting there are $N = 500$ marginal densities for each of $\mu^*$ and $\sigma^{2, *}$, the plotted  densities for each parameter are weighted according to the proportion of vehicles in a sample of $M=100$ draws of $(\bm \theta^*_{1:N}, \bm k_{1:N})$ obtained from the UVB approximation. That is, the weights are calculated according to
\begin{equation}
	w_j = \sum_{m=1}^{M} \sum_{i=1}^N \frac{I(k_{i}^{(m)} = j)}{MN},
\end{equation}
so that $w_j$ represents the proportion of the $M N$ many sampled $k_i$ values, denoted by $k_{i}^{(m)}$ for $i=1,2,...,N$ and $m=1,2,...,M$, that correspond to the given value of $j$. The weights suggest that only six of the $\bm \theta_j^*$ values account for the majority of the vehicles, with the six weighted densities associated with $\mu^*_k$ and $\sigma^*_k$ most prominent in the figures shown in panel (a). In contrast, the sample of $\bm \theta_j^*$ values that are seldom (if ever) allocated to a vehicle and hence receive little or no weight appear in these figures as flat lines indistinguishable from zero. 
\\

Now turning to panel (b) of Figure \ref{fig:DPPresults}, a predictive distribution for new values of $y$ is estimated for each cluster location $j$, using the $M$ previously simulated values $\bm \theta_{1:N}^*$. The mean of each predictive distribution is plotted against the corresponding predictive standard deviation, with the size of each point given by $w_j$. The fifty pairs of means and standard deviations shown correspond to 80\% of all simulated $k_i$ values, with the results showing that the majority of vehicles belong to a relatively small number of large and cohesive groups, each associated with a distinct predictive mean value coupled with low predictive standard deviation. Members of these groups appear to stay in the same region of their lane, but with these regions spread across both sides of the centre line. There are also many smaller groups, having predictive means closer to zero but with larger standard deviations, perhaps describing idiosyncratic vehicle positioning in the region of the centre lane. 
\\

The bottom panel plots, in grey, the individual predictive densities associated with fifty randomly selected vehicles, with the average predictive density over all $N=500$ vehicles in the sample shown in dark blue. Note that the predictive distribution associated with an individual vehicle will typically itself be comprised of a mixture of components. Importantly, many of the individual predictive densities display reduced uncertainty, relative to the overall average.
\\

Using data up to each time period $T_n$, we predict the future position $y_{i, T_n+h}, h = 1, 2, \ldots, 50$ for each vehicle using four different predictive distributions described in Section~\ref{sec:UVBpredicting}:
\begin{enumerate}
\item The DPM predictive distribution (\ref{predDPM}), with approximate inference provided via UVB.
\item The DPM predictive distribution (\ref{predDPM}), with approximate inference provided via MFVB,
\item The DPM predictive distribution (\ref{predDPM}), with approximate inference provided via SVB,
\item The independent model predictive distribution (\ref{predT}), with exact inference.
\end{enumerate}

\begin{figure}[h]
\centering
\includegraphics[width=0.95\textwidth]{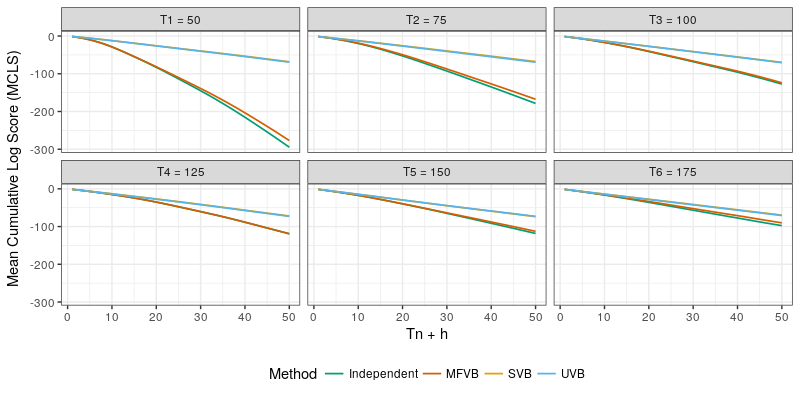}
\caption{Mean cumulative predictive log scores (MCLS) for each model averaged across $N=500$ vehicles. Each model is fit using data up to $T_n$, then forecasts are made for each of the following fifty observations. The SVB and UVB implementations are nearly identical, while the MFVB implementation performs only slightly better than the fully independent model.}
\label{fig:dppForecast}
\end{figure}

The mean cumulative predictive log scores (MCLS), averaged across each of the $N=500$ vehicles, and associated with each of the four types of predictive distributions for individual cars enumerated above, are plotted in Figure \ref{fig:dppForecast}.
\\

The results show that, while in each case both approximate implementations of the DPM model outperform the analytically exact independent model, the posterior dependency in the SVB and UVB approximations greatly improves forecasts relative to MFVB. The UVB and SVB lines coincide, and there is no evidence of accumlating approximation error through the UVB recursion relative to the single model fit of SVB. As the amount of data increases, the MFVB and independent model log scores similarly increase. In contrast, the UVB inference MCLS stays at the same level: the $N \times (T_6 - T_1) = 62,500$ additional observations included in $T_6$ has not provided much marginal information to improve forecasts relative to the original $T_1$ fit with $N \times T_1 = 25,000$ observations. By construction the DPM shares information between vehicles, so forecasts of vehicle $i$ are accurate even with only $T_1 = 50$ observations of that particular vehicle. When MFVB inference is employed forecasts are only slightly stronger than the fully independent model that does not share information, implying that the MFVB implementation did not successfully include behaviour of other vehicles.
\\

\section{Conclusions}
\label{sec:UVBSummary}

This paper proposes a framework to extend the use of SVB inference to a sequential posterior updating setting. UVB is a variational analogue to exact Bayesian updating, where the previous posterior distribution, taken as the prior for the update, is replaced with an approximation itself derived from an earlier SVB approximation. This allows a sequence of posterior distributions to be available in an online data setting, before observation of the complete dataset. Repeated updates may incur a large computational overhead from multiple stochastic optimisation algorithms, and so UVB-IS is proposed as an extension to UVB, as it significantly reduces the computational load of updates by using draws from previous posterior approximations, as well as their associated log-likelihoods, in an importance sampler. 
\\

We demonstrate the application of UVB and UVB-IS in two simulation settings, namely for time series forecasting and clustering, as well as in the empirical `Eight Schools' example involving a hierarchical model. In each case the approximation error introduced by UVB and UVB-IS is assessed relative to both exact inference, and a standard implementation of SVB, where all past observations are used in a single optimisation. We find that, despite UVB introducing additional approximations, there are mixed results in terms of accuracy and computational speed between UVB and SVB, with SVB having less error and being faster in scenarios where the log-likelihood component of the gradient estimation is small. However, in the clustering model, where the log-likelihood computation dominates the gradient estimation, UVB outperforms SVB in terms of both accuracy and speed. In each case UVB-IS introduces a small amount of additional error relative to UVB, but nevertheless still achieves large computational gains. In addition, the variance of the gradient estimator in UVB and UVB-IS is shown to be orders of magnitude less than that of SVB, which may partially offset the approximation error in the prior used by the updating methods.
\\

The proposed UVB and UVB-IS algorithms are highly suited to large time-series data streams where up-to-date inference is required at all times to inform decisions, including those based on forecasts and where data arrives so rapidly as to render MCMC infeasible. To illustrate this type of situation, an empirical illustration regarding observed lane positions of vehicles on the US-101 Highway is presented using a Dirichlet Process Mixture. In this implementation of UVB, an approximating distributional family that exploits dependence between cluster locations and indicator variables is detailed. Forecasts of future lane positions produced using UVB are comparable to an SVB approach. Posterior dependence is induced by exploiting the known full conditional distribution for the discrete indicator variables by using these as a component of the approximating distribution. Infering the model through UVB and SVB outperform inference using MFVB, as this method requires an independent posterior approximation. Future research involves the application of UVB to build a more sophisticated heterogeneous model to provide forecasts of vehicle movement from this dataset in an online fashion -- where UVB facilitates model updates and forecasts in a short time-frame after data arrives.\\
		 
\bibliographystyle{chicago}
\bibliography{references}

\appendix

\section{Calculation of Lateral Lane Deviation}\label{AppA}

\begin{figure}[htbp]
\centering
\includegraphics[width = 0.8\textwidth]{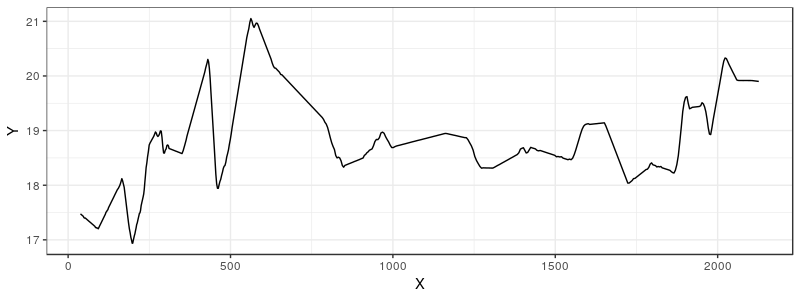}
\caption{Coordinate system for one vehicle. The $X-$axis denotes distance travelled along the lane, and the $Y-$axis denotes the relative vertical position in the lane.}
\label{fig:coords}
\end{figure}

Let $x_{i, t}$ denote the position of vehicle $i$ along the direction of travel at time $t$, and $y_{i, t}$ denote the position across the lane, as in Figure \ref{fig:coords} for one vehicle travelling to the right. 

\begin{figure}[htbp]
\centering
\includegraphics[width=0.65\textwidth]{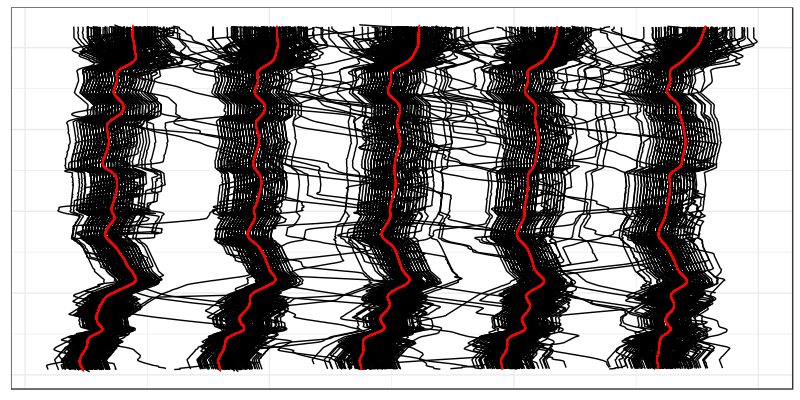}
\caption{The fitted spline models for each lane in red overlaid on the raw data in black.}
\label{fig:transform}
\end{figure}

For each vehicle $i$ and time $t$ since entering the road, with travel originating at $t = 1$, the total distance travelled up to time $t$ is given by 
\begin{equation}
\label{transform:distance}
d_{i, t} = \sum_{s=2}^t \sqrt{(x_{i, s} - x_{i, s-1})^2 + (y_{i, s} - y_{i, s-1})^2}.
\end{equation}
Using this distance measure and 100 randomly sampled vehicles per lane, the two-dimensional coordinates corresponding to the centre line of each lane are estimated via independent smoothing splines, where each coordinate is a function of the distance travelled to that point. Each smoothing spline is calculated using the `R stats' package \citep{R}. The estimated centre line for lane $k$, is denoted by the curve $\{\widehat{x}_{d,k} = f_{x}^k(d), \widehat{y}_{d,k} = f_{y}^k(d)\}$, for $d \geq 0$. The fitted spline models are shown in red overlaying the raw data in Figure \ref{fig:transform}
\\

Excluding the vehicles used to estimate the spline models, each of the vehicles in the dataset uses the relevant lane centre line estimate fit from the spline model associated with its starting lane to calculate relative coordinates $\{x^*_{i, t}, y^*_{i, t}\}$. 
$x^*_{i, t}$ denotes the distance travelled along the road, and $y^*_{i, t}$ denotes the deviation from the lane centre line, and are calculated by
\begin{align}
\widehat{d}_{i, t} &= \arg \underset{d}{\min} \sqrt{(x_{i, t} - f_x^k(d))^2 + (y_{i, t} - f_y^k(d))^2}\\
\widehat{x}_{i, t} &= f_x^k(\widehat{d}_{i, t}) \\
\widehat{y}_{i, t} &= f_y^k(\widehat{d}_{i, t}) \\
y^*_{i, t} &= \mbox{sign}\left(\tan\left(\frac{\widehat{x}_{i, t} - x_{i, t}}{\widehat{y}_{i, t} - y_{i, t}} \right)^{-1} - \tan \left(\frac{f_x^{\prime, k}(\widehat{d}_{i, t})}{f_y^{\prime, k}(\widehat{d}_{i, t})}\right)^{-1} \right) \sqrt{(x_{i, t} - \widehat{x}_{i, t})^2 + (y_{i, t} - \widehat{y}_{i, t})^2} \label{transform:xstar}
\end{align}
The coordinate pair $(\widehat{x}_{i, t}, \widehat{y}_{i, t})$ denotes the closest position of the spline model to the actual vehicle position given by the pair $(x_{i, t}, y_{i, t})$. Lateral deviation $y^*_{i, t}$ has magnitude equal to that of the vector from $(\widehat{x}_{i, t}, \widehat{y}_{i, t})$ to $(x_{i, t}, y_{i, t})$. A negative sign on $y^*_{i, t}$ indicates that the vehicle is to the left of the lane centre, and a positive sign indicates that the vehicle is to the right of the lane centre.

\section{Equivalence of augmented and marginal KL Divergence gradients}\label{AppB}

Consider the augmented posterior distribution
\begin{equation}
\label{appendixPost}
p(\theta, k | y) \propto p(y | \theta, k)p(k| \theta) p(\theta)
\end{equation}
and variational approximation given by
\begin{equation}
\label{appendixApprox}
q_{\lambda}(\theta, k | y) = q_{\lambda}(\theta| y)p(k | y, \theta).
\end{equation}
The corresponding KL divergence, $KL[q_{\lambda}(\theta, k | y)\hspace{.1cm}||\hspace{.1cm}p( \theta, k | y)]$, is indirectly minimised using the gradient
\begin{equation}
\label{KLgradappendix}
\frac{\partial KL[q_{\lambda}(\theta, k | y)\hspace{.1cm}||\hspace{.1cm}p( \theta, k | y)]}{\partial \lambda} = -\frac{\partial \mathcal{L}(q, \lambda)}{\partial \lambda},
\end{equation}
where the gradient $\partial \mathcal{L}(q, \lambda) / \partial \lambda$ is the score-based gradient of the ELBO, given by
\begin{equation}
\label{elbogradappendix}
\frac{\partial \mathcal{L}(q, \lambda)}{\partial \lambda} =
\int_{\theta, k} q_{\lambda}(\theta, k | y) \frac{\partial \log(q_{\lambda}(\theta, k | y))}{\partial \lambda} \left(\log(p(y, \theta, k)) - \log(q_{\lambda}(\theta, k | y)\right) d \theta d k.
\end{equation}
Next, consider the associated marginal posterior distribution,
\begin{equation}
p(\theta | y) \propto p(y | \theta) p(\theta),
\end{equation}
and consider using as the variational approximation $\widetilde{q}_{\lambda}(\theta|y)$ given by the first component (only) on the right hand side of (\ref{appendixApprox}), i.e. $\widetilde{q}_{\lambda}(\theta|y) \equiv q_{\lambda}(\theta|y)$. Note that, as a consequence, $\log{q_{\lambda}} \equiv \log{\widetilde{q}_{\lambda}}$ and 
$\frac{\partial \log{q_{\lambda}}}{\partial \lambda} \equiv \frac{\partial \log{\widetilde{q}_{\lambda}}}{\partial \lambda}$.
The KL divergence in this case, $KL[\widetilde{q}_{\lambda}(\theta | y)\hspace{.1cm}||\hspace{.1cm}p(\theta | y)]$, has gradient given by
\begin{equation}
\label{KLgradappendix2}
\frac{\partial KL[\widetilde{q}_{\lambda}(\theta | y)\hspace{.1cm}||\hspace{.1cm}p(\theta | y)]}{\partial \lambda} = -\frac{\partial \mathcal{L}(\widetilde{q}, \lambda)}{\partial \lambda},
\end{equation}
where $\partial \mathcal{L}(\widetilde{q}, \lambda) / \partial \lambda$ is
\begin{equation}
\label{elbogradappendix2}
\frac{\partial \mathcal{L}(\widetilde{q}, \lambda)}{\partial \lambda} = 
\int_{\theta} \widetilde{q}_{\lambda}(\theta| y) \frac{\partial \log(\widetilde{q}_{\lambda}(\theta | y))}{\partial \lambda} \left( \log(p(y, \theta)) - \log(\widetilde{q}_{\lambda}(\theta | y))\right) d\theta.
\end{equation}
Here we show that (\ref{elbogradappendix}) is equal to (\ref{elbogradappendix2}), and hence the gradient of both KL divergences are equal, and must share local minima.
\\

Begin by expanding each joint density in (\ref{elbogradappendix}) by (\ref{appendixPost}) and (\ref{appendixApprox}),
\begin{align}
\frac{\partial \mathcal{L}(q, \lambda)}{\partial \lambda} &=
\int_{\theta, k} q_{\lambda}(\theta| y)p(k | \theta, y) \frac{\partial( \log(q_{\lambda}(\theta | y)) + \log(p(k | \theta, y))}{\partial \lambda} \nonumber \\
&\times \left(\log(p(\theta)p(y |k, \theta)p(k | \theta)) - \log(q_{\lambda}(\theta | y)p(k | \theta, y))\right) d \theta d k \\
&= \int_{\theta, k} q_{\lambda}(\theta| y)p(k | \theta, y) \frac{\partial \log(q_{\lambda}(\theta | y))}{\partial \lambda} \nonumber \\
&\times \left(\log(p(\theta) + \log\left(\frac{p(y |k, \theta)p(k | \theta))p(y | \theta)}{p(y|\theta)}\right) - \log(q_{\lambda}(\theta | y)) - \log(p(k | \theta, y))\right) d \theta d k
\end{align}
as the term $\log(p(k | \theta, y))$ is independent of $\lambda$. Then
\begin{align}
\frac{\partial \mathcal{L}(q, \lambda)}{\partial \lambda} &= \int_{\theta, k} q_{\lambda}(\theta| y)p(k | \theta, y) \frac{\partial \log(q_{\lambda}(\theta | y))}{\partial \lambda} \nonumber \\
&\times \left(\log(p(\theta) + \log(p(y | \theta)) + \log(p(k |y, \theta)) - \log(q_{\lambda}(\theta | y)) - \log(p(k | \theta, y))\right) d \theta d k
\end{align}
by Bayes' Rule. Cancelling $\log(p(k |y, \theta))$ results in
\begin{align}
\frac{\partial \mathcal{L}(q, \lambda)}{\partial \lambda} &= \int_{\theta, k} q_{\lambda}(\theta| y)p(k | \theta, y) \frac{\partial \log(q_{\lambda}(\theta | y))}{\partial \lambda} \left(\log(p(\theta) + \log(p(y | \theta)) - \log(q_{\lambda}(\theta | y))\right) d \theta d k \\
&= \int_{\theta} \left(\int_{k}p(k | \theta, y)dk \right)\hspace{1mm} q_{\lambda}(\theta| y) \frac{\partial \log(q_{\lambda}(\theta | y))}{\partial \lambda} \left(\log(p(y, \theta) - \log(q_{\lambda}(\theta | y))\right) d \theta \label{appendixfinal}.
\end{align}
The final expression (\ref{appendixfinal}) is equivalent to the marginal model gradient (\ref{elbogradappendix2}) and the proof is complete.

\section{Mean Field Variational Bayes implementation of the Dirichlet Process Mixture}\label{AppC}

Implementation of MFVB for this model follows the offline coordinate ascent approach of \cite{Wang2011}, employing the stick-breaking construction of the Dirichlet Process as
\begin{align}
\theta_{j}^* &\overset{iid}{\sim} G_0, \\
\beta_j^{\prime} &\overset{iid}{\sim} Beta(1, \alpha), \\
\beta_N^{\prime} &= 1, \\
\beta_j &= \beta_j^{\prime} \prod_{l=1}^{j-1}(1 - \beta_l^{\prime}), \\
G &= \sum_{j=1}^N \beta_j \delta(\theta_j^*),
\end{align}
where $\delta$ is the Dirac Delta function. The stick-breaking construction is equivalent to the CRP representation of the DP, after marginalisation over $\beta$ \citep{Miller2018}, and is similarly augmented with the set of indicator variables $k_{1:N}$. In this case the prior distribution is given by
\begin{equation}
k_i \sim Multinomial(\bm \beta_{1:N}).
\end{equation}
The contribution to the likelihood from observation $i$ is then determined by
\begin{equation}
\bm y_{i, 1:T_1} | \bm \theta^*_{1:N}, k_i \sim \mathcal{N}\left(\mu^*_{k_i}, \sigma_{k_i}^{2*}\right).
\end{equation}
To maintain the analytical tractablility of the MFVB approximation, we replace the base distribution $G_0$ with a conjugate prior for the normal likelihood,
\begin{align}
\mu^* | G_0 &\sim \mathcal{N}(0, 10) \\
\sigma^{2*} | G_0 &\sim Inverse Gamma(shape = \alpha_0, scale = \kappa_0)
\end{align}
where $\alpha_0$ and $\kappa_0$ are chosen to be the MLE values for the inverse gamma distribution, estimated from 100,000 samples of the implied lognormal$(0, 10)$ distribution for $\sigma^{2*}$ that was used in the SVB and UVB approaches. These values are estimated by the second algorithm of \cite{Llera2016} as
\begin{align}
\alpha_0 &= 0.15275 \\
\kappa_0 &= 0.00102.
\end{align}

The variational approximation employed is of the form
\begin{equation}
q_{\bm \lambda_n}(\bm k_{1:N}, \bm \beta_{1:N}^{\prime}, \bm \theta^*_{1:N}) = \prod_{i=1}^N q(k_i)q(\beta_i^{\prime})q(\mu_i^*)q(\sigma^{2*}_i)
\end{equation}
with
\begin{align}
k_i &\sim Multinom(\bm \rho_i) \\
\beta^{\prime}_i &\sim Beta(a_i, b_i) \\
\mu_i^* &\sim \mathcal{N}(\gamma_i, \tau_i) \\
\sigma_i^{2*} &\sim Inv.Gamma(\alpha_i, \kappa_i)
\end{align}
Coordinate ascent algorithms for MFVB consists of cycling through in a set of equations for each parameter until the change in each parameter is below some threshold. For this model the equations are given by
\begin{align}
a_j &= 1 + \sum_{i=1}^N \rho_{ij}, \\
b_j &= \alpha + \sum_{i=1}^N \sum_{l=j+1}^N \rho_{il}, \\
\rho_{ij} &\propto -\frac{T_n}{2} E_q[\log(\sigma^{2*}_j)] - \frac{\alpha_j}{2 \kappa_j} \left(\sum_{t=1}^{T_n} y_{it}^2 - 2 \gamma_j \sum_{t=1}^{T_n}y_{it} + T_n (\gamma_j^2 + \tau_j) \right) + E_q[\log(\beta_j)], \\
\gamma_j &= \frac{10 \frac{\alpha_j}{\kappa_j} \sum_{i=1}^N \sum_{t=1}^{T_n}\rho_{ij} y_{it}}{10 T_n\frac{\alpha_j}{\kappa_j} \sum_{i=1}^N \rho_{ij} +1}, \\
\tau_j &= \frac{10}{10 T_n\frac{\alpha_j}{\kappa_j} \sum_{i=1}^N \rho_{ij} +1}, \\
\alpha_j &= \alpha_0 + \frac{T_n}{2} \sum_{i=1}^N \rho_{ij}, \\
\kappa_j &= \kappa_0 + \frac{1}{2} \left(\sum_{i=1}^N \rho_{ij}\left(\sum_{t=1}^{T_n}y_{it}^2 + T_n(\gamma_j^2 + \tau_j) - 2 \gamma_j \sum_{t=1}^{T_n}y_{it} \right) \right).
\end{align}
The expectations $E_q[\log(\beta_j)]$ are available in closed form as
\begin{equation}
E_q[\log(\beta_j)] = E_q[\log(\beta_j^{\prime})] + \sum_{l=1}^{j-1}E_q[\log(1 - \beta_j^{\prime})]
\end{equation}
where
\begin{equation}
E_q[\log(\beta_j^{\prime})] = \Psi(a_j) - \Psi(a_j + b_j)
\end{equation}
and
\begin{equation}
E_q[\log(1 - \beta_j^{\prime})] = \Psi(b_j) - \Psi(a_j + b_j)
\end{equation}
where $\Psi$ is the digamma function. The expectation $E_q[\log(\sigma^{2*}_j)]$ does not have a closed form solution but is estimated from samples of $q(\sigma^{2*}_j)$.

\end{document}